\PassOptionsToPackage{unicode}{hyperref}
\PassOptionsToPackage{hyphens}{url}
\PassOptionsToPackage{dvipsnames,svgnames,x11names}{xcolor}
\documentclass[12pt]{article}

\usepackage{amsmath,amssymb}
\usepackage{enumitem}
\usepackage{iftex}
\ifPDFTeX
  \usepackage[T1]{fontenc}
  \usepackage[utf8]{inputenc}
  \usepackage{textcomp} 
\else 
  \usepackage{unicode-math}
  \defaultfontfeatures{Scale=MatchLowercase}
  \defaultfontfeatures[\rmfamily]{Ligatures=TeX,Scale=1}
\fi
\usepackage{lmodern}
\ifPDFTeX\else  
\fi
\IfFileExists{upquote.sty}{\usepackage{upquote}}{}
\IfFileExists{microtype.sty}{
  \usepackage[]{microtype}
  \UseMicrotypeSet[protrusion]{basicmath} 
}{}
\makeatletter
\@ifundefined{KOMAClassName}{
  \IfFileExists{parskip.sty}{%
    \usepackage{parskip}
  }{
    \setlength{\parindent}{0pt}
    \setlength{\parskip}{6pt plus 2pt minus 1pt}}
}{
  \KOMAoptions{parskip=half}}
\makeatother
\usepackage{xcolor}
\makeatletter
\ifx\paragraph\undefined\else
  \let\oldparagraph\paragraph
  \renewcommand{\paragraph}{
    \@ifstar
      \xxxParagraphStar
      \xxxParagraphNoStar
  }
  \newcommand{\xxxParagraphStar}[1]{\oldparagraph*{#1}\mbox{}}
  \newcommand{\xxxParagraphNoStar}[1]{\oldparagraph{#1}\mbox{}}
\fi
\ifx\subparagraph\undefined\else
  \let\oldsubparagraph\subparagraph
  \renewcommand{\subparagraph}{
    \@ifstar
      \xxxSubParagraphStar
      \xxxSubParagraphNoStar
  }
  \newcommand{\xxxSubParagraphStar}[1]{\oldsubparagraph*{#1}\mbox{}}
  \newcommand{\xxxSubParagraphNoStar}[1]{\oldsubparagraph{#1}\mbox{}}
\fi
\makeatother

\usepackage{longtable,booktabs,array}
\usepackage{calc} 
\usepackage{etoolbox}
\makeatletter
\patchcmd\longtable{\par}{\if@noskipsec\mbox{}\fi\par}{}{}
\makeatother
\IfFileExists{footnotehyper.sty}{\usepackage{footnotehyper}}{\usepackage{footnote}}
\makesavenoteenv{longtable}
\usepackage{graphicx}
\makeatletter
\def\maxwidth{\ifdim\Gin@nat@width>\linewidth\linewidth\else\Gin@nat@width\fi}
\def\maxheight{\ifdim\Gin@nat@height>\textheight\textheight\else\Gin@nat@height\fi}
\makeatother
\setkeys{Gin}{width=\maxwidth,height=\maxheight,keepaspectratio}
\makeatletter
\def\fps@figure{htbp}
\makeatother

\makeatletter
\@ifpackageloaded{caption}{}{\usepackage{caption}}
\AtBeginDocument{%
\ifdefined\contentsname
  \renewcommand*\contentsname{Table of contents}
\else
  \newcommand\contentsname{Table of contents}
\fi
\ifdefined\listfigurename
  \renewcommand*\listfigurename{List of Figures}
\else
  \newcommand\listfigurename{List of Figures}
\fi
\ifdefined\listtablename
  \renewcommand*\listtablename{List of Tables}
\else
  \newcommand\listtablename{List of Tables}
\fi
\ifdefined\figurename
  \renewcommand*\figurename{Figure}
\else
  \newcommand\figurename{Figure}
\fi
\ifdefined\tablename
  \renewcommand*\tablename{Table}
\else
  \newcommand\tablename{Table}
\fi
}
\@ifpackageloaded{float}{}{\usepackage{float}}
\floatstyle{ruled}
\@ifundefined{c@chapter}{\newfloat{codelisting}{h}{lop}}{\newfloat{codelisting}{h}{lop}[chapter]}
\floatname{codelisting}{Listing}

\makeatother
\makeatletter
\@ifpackageloaded{caption}{}{\usepackage{caption}}
\@ifpackageloaded{subcaption}{}{\usepackage{subcaption}}
\makeatother

\ifLuaTeX
  \usepackage{selnolig}  
\fi
\usepackage[authoryear]{natbib}
\usepackage{bookmark}

\IfFileExists{xurl.sty}{\usepackage{xurl}}{} 
\hypersetup{
  pdftitle={Title},
  pdfauthor={Author 1; Author 2},
  pdfkeywords={3 to 6 keywords, that do not appear in the title},
  colorlinks=true,
  linkcolor={blue},
  filecolor={Maroon},
  citecolor={Blue},
  urlcolor={Blue},
  pdfcreator={LaTeX via pandoc}}

\newcommand{\anon}{1}

\usepackage{amsthm}

\usepackage{amsmath}
\DeclareMathOperator*{\argmax}{arg\,max}

\usepackage{lmodern}
\usepackage{comment}

\theoremstyle{definition}
\newtheorem{assumption}{Assumption}
\newtheorem{step}{Step}
\newtheorem{remark}{Remark}

\theoremstyle{plain}
\newtheorem{lem}{\protect\lemmaname}
\newtheorem{thm}{\protect\theoremname}
\newtheorem{algorithm}{Algorithm}

\usepackage{caption}
\usepackage{subcaption}
\usepackage{graphicx}
\usepackage{array}
\usepackage{booktabs}
\usepackage{multicol} 
\usepackage{float}
\usepackage{threeparttable}

\renewcommand{\qed}{\hfill{\tiny \ensuremath{\blacksquare} }}%

\newcommand{\Ep}{\mathrm{E}}

\newcommand{\mS}{\mathcal{S}}
\newcommand{\mY}{\mathcal{Y}}

\newcommand{\dd}{\mathrm{d}}

\renewcommand{\qed}{\hfill {\tiny {\ensuremath{\blacksquare}}}}

\usepackage{babel}
\providecommand{\lemmaname}{Lemma}
\providecommand{\theoremname}{Theorem}

\begin{document}

\def\spacingset#1{\renewcommand{\baselinestretch}%
{#1}\small\normalsize} \spacingset{1}


\if1\anon
{
  \title{\bf A Distribution Regression Approach to Censored Selection Models\thanks{
We thank Kevin Lang, Ruli Xiao, and seminar and conference participants at Boston University, Munich Econometrics Workshop 2023, Midwest Econometrics Group Conference 2024, and BC-BU Greenline Econometrics Workshop 2024 for helpful comments.
}}
  \author{Iv\'an Fern\'andez-Val\hspace{.2cm}\\
    Department of Economics, Boston University\\
    and \\
    Seoyun Hong \\
    Department of Economics, Georgia State University}
  \maketitle
} \fi

\if0\anon
{
  \bigskip
  \bigskip
  \bigskip
  \begin{center}
    {\LARGE\bf A Distribution Regression Approach to Censored Selection Models}
\end{center}
  \medskip
} \fi

\bigskip
\begin{abstract}
We develop a distribution regression model with a censored selection rule, offering a semi-parametric generalization of the Heckman selection model. Our approach applies to the entire distribution, extending beyond the mean or median, accommodates non-Gaussian error structures, and allows for heterogeneous effects of covariates on both the selection and outcome distributions. By employing a censored selection rule, our model can uncover richer selection patterns according to both outcome and selection variables, compared to the binary selection case. We analyze identification, estimation, and inference of model functionals such as sorting parameters and distributions purged of sample selection. An application to labor supply using data from the UK reveals different selection patterns into full-time and overtime work across gender, marital status, and time. Additionally, decompositions of wage distributions by gender show that selection effects contribute to a decrease in the observed gender wage gap at low quantiles and an increase in the gap at high quantiles for full-time workers. The observed gender wage gap among overtime workers is smaller, which may be driven by different selection behaviors into overtime work across genders.
\end{abstract}

\noindent%
\vfill

\newpage
\spacingset{1.8} 

\section{Introduction}

Empirical researchers frequently encounter non-random sample selection in their data.
A prominent example is the estimation of wage equations when the wages
of non-working individuals are not observed. Sample selection bias
arises when there are unobservables affecting both sample selection
and the outcome of interest, such as employment status and offered
wage. The classical Heckman selection model (HSM), introduced by \citet{heckman1974shadow}, is the most widely used method for addressing sample selection bias. It relies on parametric Gaussian error structures and assumes homogeneous effects of exogenous covariates. Recently, \citet{chernozhukov2025distribution}
(CFL) developed a distribution regression (DR) model that accounts
for endogenous sample selection in an effort to generalize the 
HSM. They provide a method to analyze the conditional distribution,
extending beyond the mean or median, while taking sample selection
into account. Furthermore, they depart from Gaussian error structures
and allow for heterogeneous effects of covariates on both the outcome
distribution and the sample selection process.

This paper develops a distribution regression model with a censored selection rule, building on the distribution regression model with
a binary selection rule proposed by CFL. Implementing a censored selection rule requires more detailed information about the sample selection mechanism compared to using a binary selection rule. In the context of wage equations, this means modeling the employment selection equation with a censored work hours equation, instead of a binary employment status equation.\footnote{We highlight that implementing a censored selection model is often possible using available data to implement a binary selection model. As noted in \citet{fernandez2021nonseparable}, researchers often dichotomize censored selection variables to employ binary selection model such as the duration of unemployment, the length of a training program, and the magnitude or duration of welfare benefit receipt.} While employing
a censored selection rule does not aid in identification - since our
model relies on the same identification assumptions as CFL - the main advantage is that it allows for the analysis of richer patterns of selection heterogeneity. First, this approach enables modeling the distribution of the selection variable, such as the distribution of work hours, which is not possible in sample selection models that rely on a binary selection rule. Second, it allows for the estimation of selection sorting (e.g., sign of selection) according to the level of the selection variable. For instance, while binary sample selection models enable estimating positive or negative selection into employment, our approach can examine selection sorting according to work hours, such as selection into full-time or overtime work.

An example where heterogeneous selection patterns, which can be revealed by a censored selection model, are of interest, is the study of the gender wage gap. Among the many factors contributing to the gender wage gap discussed by \cite{blau2017gender}, gender differences in work hours and patterns of selection into employment are particularly relevant to our model and analysis. In terms of work hours, they highlight a significant gender disparity, noting that women comprise a larger share of the part-time workforce. This higher prevalence of part-time work among women compared to men may widen the overall gender wage gap, as part-time workers typically earn lower hourly wages than full-time workers (\citet{hirsch2005part}). On the other hand, there is literature examining how long work hours affect the gender wage gap among highly skilled workers (\citet{bertrand2010dynamics}, \citet{goldin2014grand}). 
Employing a censored selection rule and estimating the conditional distribution of work hours can help better understand the characteristics of workers with both short and long hours.

There is also a vast literature documenting different selection sorting patterns into employment by gender and their impacts on the gender wage gap. For example, \citet{mulligan2008selection} find weaker evidence of convergence in the gender wage gap after accounting for sample selection. Similarly, \citet{arellano2017quantile}, \citet{maasoumi2019gender}, and \citet{chernozhukov2025distribution} address sample selection bias in the wage distribution and highlight that examining wage inequality among the employed may present a distorted picture of overall labor market disparities. As there are significant gender differences in work hours, studying selection sorting patterns by work hours or worker types could offer valuable insights into the mechanisms behind the gender wage gap.


We analyze identification, estimation, and inference methods in a
distribution regression model that corrects for sample selection bias
using a censored selection rule. The parameters of interest are features of the joint distribution of the latent outcome and selection variable. A key distinction from CFL is that the observed selection variable is discrete or continuous, rather than binary. For identification, we rely on 
exclusion restrictions. In particular, we require the existence of a binary instrumental variable (IV)
that affects the marginal distribution of the selection variable
but does not affect the marginal distribution of latent outcome or
the relationship (sorting) between the latent outcome and selection variables. 
We propose a tractable estimation algorithm consisting of three steps involving (bivariate) probit regressions with sample selection corrections. Lastly, we provide functional central limit theorems for our estimators and uniform inference methods based on multiplier bootstrap.

We apply our method to wage and work hours data in the UK. The goal
is to study selection sorting patterns across work hours (e.g., selection
into full-time and overtime work) and conduct wage decompositions
between gender by worker types (work hours). First, we find heterogeneous
selection patterns by gender, marital status, year, and worker types.
Specifically, single men exhibit positive selection into full-time
work across all wage quantiles, while married women do so only
at lower wage quantiles. Since married men and single women do not
show significant selection effects, we interpret this as indicating
that single men and married women have greater flexibility in selecting
their work hours. Both genders display negative selection into overtime
work, with the effect being stronger at the top of the wage distribution. 

Second, we decompose the difference between the male and female wage
distributions into composition, wage structure, hours structure,
and sorting effects for full-time and over-time workers.
This is a distributional generalization of the Oaxaca-Blinder decomposition
(\citet{oaxaca1973male}, \citet{blinder1973wage})
which accounts for sample selection. The decomposition
for full-time workers reveals that aggregated hours and sorting effects narrow the observed gender wage gap at the bottom but widen it at the top of the distribution. Although the observed gap among overtime workers is smaller than among employed or full-time workers, the decomposition suggests that this reflects gender differences in hours structure and sorting behaviors. Consistently, the decomposition of work hours shows that the structure effect explains most of the gender differences relative to the composition effect. These findings underscore the
importance of accounting for selection and flexible heterogeneity in analyses of the gender wage gap.

\paragraph*{\textbf{Literature Review.}} The distribution regression model is a semiparametric generalization of the tobit
type-3 model and is a variant of the \citet{heckman1979sample} selection model. In \citet{amemiya1979estimation}, the tobit
type-3 model was
initially examined in a fully parametric setting, imposing additivity and
normality, and estimated by maximum likelihood.
\citet{vella1993simple} provided a two-step estimator based on estimating the
generalized residual from the selection equation and including it as a
control function in the outcome equation. \citet{honore1997estimation}, \citet{chen1997semiparametric} and \citet{lee2006semi} provide semi-parametric estimators for this
model. 

\citet{fernandez2021nonseparable} showed nonparametric identification of local average and distributional effects in the selected population conditional on a control variable, the conditional distribution of the selection variable, even without exclusion restrictions. To identify unconditional effects over the entire population, they relied on exclusion restrictions for an instrumental variable with large support. We focus on unconditional objects without rich instrumental variation by restricting local dependence between the outcome and selection variable. \citet{chen2024estimation} adopted a quantile regression model with a censored selection rule, building on the binary selection model of \citet{arellano2017quantile}. They do not rely on exclusion restrictions, but impose parametric structure on the sorting between the latent outcome and selection variables. In contrast, our exclusion restrictions impose semiparametric structure on the sorting. Thus, while our sorting parameter is infinite dimensional, the sorting parameter of \citet{chen2024estimation} is finite dimensional. Moreover, the quantile regression approach relies on continuity of the outcomes, whereas our distribution regression approach can accommodate continuous, discrete, and mixed outcomes.

Recently, \cite{chen2025distribution} proposed a distribution regression model with sample selection that also incorporates a censored selection rule. However, their identification and estimation frameworks differ from ours. Our identification framework is a localized extension of CFL. We retain their outcome exclusion, while requiring the relevance condition for the marginal distribution and the exclusion restriction on selection sorting to hold at a single reference value of the selection variable.
When this reference value is the censoring point, our identifying assumptions coincide with those of CFL.\footnote{Note that CFL impose the exclusion restriction on selection sorting only at the threshold value of zero for the latent selection variable, which is the only relevant value in the binary selection setting; see Assumption 1 in CFL. This contrasts with Remark 1 of \cite{chen2025distribution}, which states that CFL impose the exclusion restriction at all values of the latent selection variable.} 
The censored selection rule therefore allows us to identify richer parameters in the selection and selection sorting equations than in the binary selection setting, without imposing assumptions beyond those required by CFL. By contrast, \cite{chen2025distribution} relaxed CFL's outcome exclusion restriction by exploiting the richer information provided by the non-binary selection variable. This relaxation is accompanied by a stronger selection sorting exclusion restriction than in our framework, as their restriction must hold globally. 
Therefore, the two approaches are complementary, as they utilize the additional information provided by the censored selection through different identification strategies while ultimately identifying the same parameters. We focus on identifying additional selection-related parameters while maintaining CFL's identification framework, while \cite{chen2025distribution} use censored selection to relax part of CFL's identification assumptions. We discuss the differences in identification and estimation procedures in detail in the main text. 

We also contribute to the literature on distribution regression with endogenous regressors. \cite{briseno2020flexible} develop a generalized additive model for location, scale and shape (GAMLSS) that accommodates endogenous regressors using two-stage residual inclusion estimation, a form of the control function approach. 
 \cite{wied2024semiparametric} develops an instrumental variable estimator for a semiparametric DR model that employs a control function approach and extends IV probit estimation. 

\paragraph*{\textbf{Outline.}} Section \ref{sec:identification} establishes identification under censored sample selection using a local Gaussian representation. Section \ref{sec:dr} develops the distribution regression model and presents estimation and inference results. Section \ref{sec:empirics} reports the empirical application, and Section \ref{sec:conclusion} concludes. The supplementary material (SM) contains additional discussions and results, a Monte Carlo simulation calibrated to the empirical application, and proofs.

\section{Identification  under Censored Selection}\label{sec:identification}

We first recall the local Gaussian representation (LGR) of CFL. We use this representation to analyze the identification of the joint distribution of the latent outcome and selection variables under a censored selection rule.

\subsection{Local Gaussian Representation}
Let $(S^{*},Y^{*})$ be two random variables with joint cumulative
distribution function (CDF) $F_{S^{*},Y^{*}}$ and marginal CDFs $F_{S^{*}}$
and $F_{Y^{*}}$. In the sample selection model, $Y^*$ will be the latent outcome of interest and $S^*$ the latent selection variable that determines when $Y^*$ is observed. The following lemma from CFL shows that $F_{S^{*},Y^{*}}$ has a representation at each point as a bivariate standard Gaussian distribution with local correlation parameter.
\begin{lem}[LGR]\label{lemma:lgr}
  $F_{S^{*},Y^{*}}$ can be represented as
\[
F_{S^{*},Y^{*}}(s,y)=P(S^{*}\leq s,Y^{*}\leq y)\equiv\Phi_{2}\left(\Phi^{-1}(F_{S^{*}}(s)),\Phi^{-1}(F_{Y^{*}}(y));\rho(s,y)\right), \ \ (s,y) \in \mathbb{R}^2,
\]
where $\Phi_{2}(\cdot,\cdot;\rho)$ is the joint CDF of a standard
bivariate Gaussian random variable with correlation parameter $\rho\in[-1,1]$. In the previous representation,  $\rho(s,y)\in[-1,1]$ is a local correlation parameter that
varies with the evaluation point $(s,y)$.
\end{lem}
Lemma \ref{lemma:lgr} establishes that $F_{S^{*},Y^{*}}$ can be represented by
a sequence of standard bivariate normal distributions with correlation parameters that depend on the evaluation point $(s,y)$. The LGR is a convenient tool to analyze identification of the sample selection model because it connects naturally with the bivariate Gaussian distribution, which is the basis of the HSM. Thus, the local correlation parameter of the bivariate Gaussian distribution is constant and equal to the correlation coefficient, that is, $\rho(s,y) = \rho$ for all $(s,y)$.
More generally, the local correlation parameter $\rho(s,y)$ is a measure of local dependence.\footnote{In particular, $\rho(s,y)=0$
if and only if $F_{S^{*},Y^{*}}(s,y)= F_{S^{*}}(s) F_{Y^{*}}(y)$, i.e., $S^*$ and $Y^*$ are locally independent at $(s,y)$. Furthermore, $\rho(s,y)$ is positive if and only if the correlation between $1\left(S^{*}\leq s\right)$ and $1\left(Y^{*}\leq y\right)$ is positive.} In the sample selection model, we will refer to $\rho(s,y)$ as the sorting parameter because it determines the local sign and extent of the selection.

It is convenient to introduce the notations for the parameters of the LGR associated with the marginal distributions. Let $\mu(s) := \Phi^{-1}(F_{S^{*}}(s))$ and $\nu(y) := \Phi^{-1}(F_{Y^{*}}(y))$ where $\mu(s)\in\bar{\mathbb{R}}$, $\nu(y)\in\bar{\mathbb{R}}$
and $\bar{\mathbb{R}}=\mathbb{R}\cup\left\{ -\infty,\infty\right\} $
is the extended real number line. Thus, we can express the LGR of $F_{S^{*},Y^{*}}$
at point $(s,y)$ as $F_{S^{*},Y^{*}}(s,y)\equiv\Phi_{2}(\mu(s),\nu(y);\rho(s,y))$. We note that the representation can be easily extended to 
CDFs conditional on covariates by allowing all parameters to depend on the value
of covariates, which we use later in the identification.

Now, we introduce the sample selection problem with a censored selection rule. The observed variables $(S,Y)$ can be defined in terms of the
latent variables $(S^{*},Y^{*})$. That is, we observe the random variables
$S$ and $Y$ such that
$S=\max(S^{*},0)$ and $Y = Y^{*}$ if $S>0$. Note that the sample selection is modeled through a censored selection
rule with a maximum function rather than using a binary selection
model with an indicator function.\footnote{Other censoring points and right censoring reduce to the baseline case by recentering and sign reversal. If $S=\max(S^{*},\underline{s})$ for $\underline{s}\neq 0$, define $S^{**} := S^*-\underline{s}$ and $\tilde S = S - \underline{s}$ such that  $\tilde S = \max(S^{**},0)$. If $S = \min(S^*,0)$, define $S^{**} := - S^*$ and $\tilde S = -S$, such that $\tilde S = \max(S^{**},0)$.} For example, if $S^*$ is desired work hours and $Y^*$ is offered wage, the observed work hours are censored at $0$ and we only observe the offered wage for those who work a positive number of hours.

\subsection{Identification via Exclusion Restrictions}
The goal is to determine if we can identify the parameters of the LGR of the latent variables $(S^*,Y^*)$, $(\mu(s),\nu(y),\rho(s,y))$,
from the joint distribution of the observed variables $(S,Y)$.  We first show that $\nu(y)$ and $\rho(s,y)$ are partially identified without
further assumptions. Then we establish point identification using exclusion restrictions. For any $s_{1}>s_{0}\geq0$, we can write the distribution of observed
variables as
\begin{eqnarray}
\Pr(S\leq s_{i}) & = & \Phi(\mu(s_{i})),\quad i\in\{0,1\}, \label{eq:sel-probs} \\
\Pr(S>s_{0},Y\leq y) & = & \Phi(\nu(y))-\Phi_{2}(\mu(s_{0}),\nu(y);\rho(s_{0},y)),\label{eq:prob1}\\
\Pr(s_{0}<S\leq s_{1},Y\leq y) & = & \Phi_{2}(\mu(s_{1}),\nu(y);\rho(s_{1},y))-\Phi_{2}(\mu(s_{0}),\nu(y);\rho(s_{0},y)) \label{eq:prob2}.
\end{eqnarray}
The selection probabilities pin down $(\mu(s_{0}),\mu(s_{1}))$ in
\eqref{eq:sel-probs} as $\mu(s_{i})=\Phi^{-1}(P(S\leq s_{i}))$. However, in identifying
the three parameters $(\nu(y),\rho(s_{0},y),\rho(s_{1},y))$, there are
only two free probabilities \eqref{eq:prob1} and \eqref{eq:prob2}, which generally leads to partial identification.
Using additional values of $S$ and $Y$ does not help for identification
as they introduce at least as many unknowns as equations.

We show that identification can be achieved using exclusion restrictions.
Let $Z$ be a potential instrumental variable and $F_{S^{*},Y^{*}|Z}$
be the joint CDF of $(S^{*},Y^{*})$ conditional on $Z$. Then, $F_{S^{*},Y^{*}|Z}$
admits the LGR
$F_{S^{*},Y^{*} \mid Z}(s,y\mid z)\equiv \Phi_{2}(\mu_{z}(s),\nu_{z}(y);\rho_{z}(s,y))$,
where $\mu_{z}(s) := \Phi^{-1}(F_{S^{*} \mid Z}(s \mid z))$ and $\nu_{z}(y) := \Phi^{-1}(F_{Y^{*} \mid Z}(y \mid z))$.  We make the following assumption:
\begin{assumption}[Exclusion Restrictions]\label{ass:excl} Assume that $Z$ is binary and for some $s_{0}\geq0$:
\begin{enumerate}[label=(\arabic*)]
\item \textit{Non-degeneracy: $0<\Pr(S>s_{0})<1$ and $0<\Pr(Z=1\mid S>s_{0})<1$ }
\item \textit{Relevance: $\Pr(S>s_{0}\mid Z=0)<\Pr(S>s_{0}\mid Z=1)<1$ }
\item \textit{Outcome exclusion: $\nu_{z}(y)=\nu(y)$ for all $y\in\mathbb{R}$
and $z\in\{0,1\}$ }
\item \textit{Sorting exclusion: $\rho_{z}(s_{0},y)=\rho(s_{0},y)$ for
all $y\in\mathbb{R}$ and $z\in\{0,1\}$. }
\end{enumerate}
\end{assumption}

Assumption \ref{ass:excl} is the same exclusion restriction as in CFL when $s_{0}=0$. The case where  $Z$ is binary is the most challenging for identification as it precludes identification coming from rich variation of $Z$ . When $Z$ is not binary,
we only need Assumption \ref{ass:excl} to be satisfied for two values of $Z$. If it holds for more than two values, the model becomes overidentified,
making the exclusion restrictions testable. If $s_{0}=0$, non-degeneracy states
that sample selection occurs and that $Z$ varies within the selected
population, and relevance requires that $Z$ affects the probability of
selection. If $s_{0}>0$, both conditions apply analogously to the subpopulation with  $S > s_0$ instead of the selected population. 
Outcome exclusion is a standard exclusion restriction,
and it holds when $Y^{*}$ is independent of $Z$. 

Sorting exclusion is a local version of the copula invariance restriction from CFL and
holds if selection sorting function at $s=s_{0}$ is independent of
$Z$. We refer to CFL for more discussion on the interpretation of this condition. Note that we require the sorting exclusion condition only for
$s=s_{0}$, so that $\rho_{z}(s,y)$ can still depend on $z$ when
$s\neq s_{0}$. This implies that we can achieve  identification
of additional parameters from the censored selection model such as
$(\mu_z(s_{1}),\rho_z(s_{1},y))$, relying on the same identification
condition as in the binary selection model. The most common case is when $s_0=0$, but exclusion might also hold for other values. For example, in the labor supply, we might have variables that determine the decision of working part-time or overtime that affect neither offered wages nor sorting.

We now demonstrate how exclusion restrictions help identify all
the parameters. Let Assumption \ref{ass:excl} hold. After imposing the exclusion restriction on the LGR, the target parameters are $(\mu_{0}(s_0),\mu_{1}(s_0),\mu_{0}(s_1),\mu_{1}(s_1),\nu(y),\rho(s_0,y),\rho_0(s_1,y),\rho_1(s_1,y))$.
First, the parameters $\mu_{0}(s_0)$, $\mu_{1}(s_0)$, $\mu_{0}(s_1)$, and $\mu_{1}(s_1)$ are identified by
\[
\mu_{z}(s)=\Phi^{-1}(\Pr(S\leq s\mid Z=z)),\ s\geq0,\; \ z\in\lbrace0,1\rbrace.
\]
At $s=s_0$, we have a system of two nonlinear equations and two unknowns
\begin{equation*}
    \Pr(S>s_0,Y\leq y\mid Z=z)=\Phi_{2}(\Phi^{-1}(\Pr(S>s_0\mid Z=z)),\nu(y);-\rho(s_0,y)), \quad z\in\lbrace0,1\rbrace,
\end{equation*}
where we have used the symmetry properties of $\Phi_2$. Using the same arguments
as in CFL, it can be shown that this nonlinear system has a unique
solution for $(\nu(y),\rho(s_0,y))$ by the global univariance result of
\citet{gale1965jacobian}. Lastly, $\rho_{z}(s,y)$ for $0 \leq s\neq s_0$ and $z\in\lbrace0,1\rbrace$
is identified from
\begin{equation}\label{eq:rho_id}
\Pr(s_{0} < S\leq s,Y\leq y\mid Z=z)=\Phi_{2}(\mu_{z}(s),\nu(y);\rho_{z}(s,y)) - \Phi_{2}(\mu_{z}(s_0),\nu(y);\rho(s_0,y)),
\end{equation}
where  $\mu_{z}(s_0)$, $\mu_{z}(s)$, $\nu(y)$ and  $\rho(s_0,y)$ are
identified from the argument given above. Equation \eqref{eq:rho_id} has a unique solution because $\rho\to\Phi_{2}(\mu,\nu;\rho)$
is strictly increasing. In the theorem below, we formulate the identification
of $\rho_z(s,y)$, as the identification results for other parameters
are stated in CFL.
\begin{thm}[Identification of $\rho_z(s,y)$]\label{thm:id}
Suppose that Assumption \ref{ass:excl} holds. Then, $\rho_z(s,y)$ is
point identified as the unique solution to \eqref{eq:rho_id}, for $z\in\{0,1\}$, $0 \leq s \neq s_0$ and $y \in \mathbb{R}$.
\end{thm}

\begin{remark}[Comparison with \cite{chen2025distribution}]
 \cite{chen2025distribution} 
 do not rely on the outcome exclusion in Assumption \ref{ass:excl}(3), but use a stronger version of the sorting exclusion in Assumption \ref{ass:excl}(4) with $\rho_{z}(s,y)=\rho(s,y)$ for all $(s,y)\in\mathbb{R}^2$ and $z\in\lbrace 0,1\rbrace$. Under this sorting exclusion, they end up with the system of equations in our notation
\begin{equation*}
    \Pr(S>s,Y\leq y\mid Z=z)=\Phi_{2}(\Phi^{-1}(\Pr(S>s\mid Z=z)),\nu_{z}(y);-\rho(s,y)), \; z\in\lbrace0,1\rbrace.
\end{equation*}
To show identification, they assume that this system is globally injective to establish that it has a unique solution for the parameters. As Remark 2 in \cite{chen2025distribution} notes, this condition is difficult to verify analytically. 
We instead verify that our identification assumptions lead to a system of equations that has a unique solution without imposing any additional assumption by employing global univalence results.\qed
\end{remark}

\section{Distribution Regression with Censored Selection}\label{sec:dr}

\subsection{Model }
Let $X$ be a vector of covariates related to $Y^*$ and $S^*$, and $Z=(X,Z_{1})$, where
$Z_{1}$ is a vector of instrumental variables, the covariates that satisfy the exclusion restrictions in Assumption \ref{ass:excl}. We consider a semi-parametric version of the LGR with covariates:
\begin{equation}\label{eq:drmodel}
    F_{S^{*},Y^{*}\mid Z}(s,y\mid z)  = \Phi_{2}(\mu_{z}(s),\nu_{z}(y);\rho_{z}(s,y)) = \Phi_{2}(-z^{\prime}\mu(s),-z^{\prime}\nu(y);g(z^{\prime}\rho(s,y))),
\end{equation}
where the exclusion restrictions are imposed as $z'\nu(y) = x'\nu(y)$ and $g(z^{\prime}\rho(s_0,y)) = g(x^{\prime}\rho(s_0,y))$, where $s_0$ is the level of selection variable where the sorting exclusion on $\rho_z(s,y)$ holds.\footnote{This follows from Assumption \ref{ass:excl}(3) and Assumption \ref{ass:excl}(4).} In the empirical application in
the next Section, $Y^{*}$ is offered wage, $S^{*}$ is desired work hours, $X$ includes labor market characteristics such as education
or experience, $Z_{1}$ includes potential out-of-work income, and $s_0=0$ is the employment participation level.

The bivariate distribution regression (BDR) model \eqref{eq:drmodel} is semiparametric in that the parameters $s\mapsto\mu(s)$, $y\mapsto\nu(y)$
and $(s,y)\mapsto\rho(s,y)$ are function-valued. $u\mapsto g(u)$
is a known link function with range $[-1,1]$ such as the Fisher trasnformation $g(u)=\tanh(u)$. We refer to $-z^{\prime}\mu(s)$ as the selection
(hours) equation, $-x^{\prime}\nu(y)$ as the outcome (wage) equation,
and $g(z^{\prime}\rho(s,y))$ as the selection sorting (hours-wage sorting)
equation. Importantly, note that the covariates of the selection sorting equations are different
for $s=s_0$ and $s \neq s_0$ as the exclusion restriction only applies when
$s=s_0$. 

The parameters $\left(\mu(s),\nu(y)\right)$ can be interpreted similarly
to those in probit models. For example, $\nu(y)$ reveals the sign
of partial effects of the corresponding covariates on the conditional
distribution of the latent outcome, $F_{Y^{*} \mid Z}(y \mid z) = \Phi(-x^{\prime}\nu(y))$, such as the conditional probability that
the latent wage $Y^{*}$ is less than  or equal to $y$. Specifically,
if $X$ is continuous, $\partial F_{Y^{*} \mid Z}(y \mid z)/\partial x =-\nu(y)\phi(-x^{\prime}\nu(y))$ where $\phi$ is the standard normal probability distribution function. The parameter $\mu(s)$ plays an equivalent
role in the partial effects on the conditional distribution of the latent
selection variable, $F_{S^{*} \mid Z}(s \mid z) = \Phi(-z^{\prime}\mu(s))$, such as the probability that work hours are less
than or equal to $s$. The parameter $\rho(s,y)$ indicates
the sign of the effect of the covariates on the sorting, since $\partial g(z^{\prime}\rho(s,y))/\partial z =\rho(s,y)\dot{g}(z^{\prime}\rho(s,y))
$ where $\dot{g}(u)=\mathrm{d} g(u)/\mathrm{d} \mathrm{u} =1-g(u)^{2}>0$ if $g(u)=\tanh(u)$. 

The main parameter of interest in our empirical application is the
selection sorting function $(z,s,y) \mapsto g(z'\rho(s,y))$. For illustration, assume
that we include only a constant in selection sorting equation, i.e. $g(z'\rho(s,y)) = g(\rho(s,y)) := \tilde \rho(s,y)$. By
definition, $\tilde \rho(s,y)$ determines the conditional local dependence between
$S^{*}$ and $Y^{*}$ at the  point $(s,y)$. In the wage 
example, $y \mapsto \tilde \rho(0,y)$ can be interpreted as selection sorting into employment
at the wage level $y$. Similarly, setting $s$ to
the appropriate level corresponding to full-time or overtime work hours,  $y \mapsto \tilde \rho(s,y)$ can be interpreted as  sorting into full-time or overtime 
work at the wage level $y$. Therefore, as mentioned in the introduction,
we can study heterogeneous patterns of sorting effects according
to selection variable $S^{*}$ and outcome variable $Y^{*}$.

Now, we compare the HSM with censored selection to
our model. HSM with outcome exclusion restriction assumes
\begin{eqnarray*}
Y^{*}  =  X^{\prime}\nu+\sigma_{U}U, \quad \text{ and } \quad
S^{*}  =  Z^{\prime}\mu+\sigma_{V}V,
\end{eqnarray*}
where $(U,V)$ is standard bivariate Gaussian with correlation
$\rho$ conditional on $Z$. The joint distribution of $(S^{*},Y^{*})$ conditional on
$Z$ is therefore modeled as 
\[
F_{S^{*},Y^{*}\mid Z}(s,y\mid z)=\Phi_{2}\left(\frac{s-z^{\prime}\mu}{\sigma_{V}},\frac{y-x^{\prime}\nu}{\sigma_{U}};\rho\right).
\]
This is a special case of \eqref{eq:drmodel}
with 
\[
-z^{\prime}\mu(s)= (s-z^{\prime}\mu)/\sigma_{V},\ -z^{\prime}\nu(y)=(y-x^{\prime}\nu)/\sigma_{U},\ g(z^{\prime}\rho(s,y))=\rho.
\]
We want to highlight three key points. First, our model is a generalization of HSM with a censored selection rule. Second, HSM only allows for location effects of covariates in the selection and outcome equations, whereas our
model permits heterogeneous effects of all covariates with respect
to $(s,y)$. Finally, HSM sets the selection sorting
parameter as a constant, while we allow for the effect of covariates
and coefficients to change with $(s,y)$, representing a significant
relaxation of the selection sorting mechanism. Furthermore, in terms
of identification, the selection sorting exclusion restriction is
trivially satisfied if we set the selection sorting parameter as a
constant. The widespread use of the classical model in empirical research,
despite its stricter assumptions, suggests that our more flexible
selection sorting exclusion restriction can be similarly validated.

\subsection{Functionals}\label{sec:functional}
We can construct functionals of the BDR's parameters and marginal distributions of the covariates, which may be of interest. For example,  the marginal distributions of  $Y^{*}$ and $S^{*}$ are
\begin{align*}
F_{Y^{*}}(y) & =\int\Phi(-x^{\prime}\nu(y))\dd F_{X}(x), \ \
F_{S^{*}}(s) =\int\Phi(-z^{\prime}\mu(s))\dd F_{Z}(z),\quad s\geq0,
\end{align*}
where $F_{X}$ and $F_{Z}$ are the CDFs of $X$ and $Z$. We can
also construct counterfactual distributions by replacing $(\nu(y),\mu(s))$
or $(F_{X},F_{Z})$ by coefficients and distributions from different
population groups. These distributions are useful to decompose the
distribution of offered wages or desired work hours between different genders, races or time periods.

We can also use the model to construct distributions for the observed
wage by worker type as defined by work hours. For $0\leq\underline{s}<\overline{s}$,
\[
F_{Y}(y\mid\underline{s}<S^{*}\leq\overline{s})=F_{Y^*}(y\mid\underline{s}<S^{*}\leq\overline{s})=\frac{F_{S^{*},Y^{*}}(\overline{s},y)-F_{S^{*},Y^{*}}(\underline{s},y)}{F_{S^{*}}(\overline{s})-F_{S^{*}}(\underline{s})},
\]
where $F_{S^{*},Y^{*}}(s,y)=\int\Phi_{2}(-z^{\prime}\mu(s),-x^{\prime}\nu(y);g(z^{\prime}\rho(s,y)))\dd F_{Z}(z)$.
We can construct corresponding counterfactual distributions by replacing $\mu(s)$, $\nu(y)$, $\rho(s,y)$, or $F_{Z}$.
In the wage application, we will decompose the differences in the
observed wage distribution between genders into changes in the worker composition
$F_{Z}$ , wage structure $\nu(y)$, hours structure $\mu(s)$, and
hours sorting $\rho(s,y)$. The hours structure and sorting effects
are new to this model, compared to the decomposition analysis in CFL.
We can construct quantiles and other functionals of the distribution
of latent and observed outcomes by applying the appropriate operator.

\subsection{Estimation}
To estimate the model parameters and functionals of interest, we assume
that we have a random sample of size $n$ from $(S,Y,Z)$, $\{(S_{i},Y_{i},Z_{i})\}_{i=1}^{n}$. In this section we will also assume that $s_0=0$, which corresponds to our empirical application. Let $(\mathcal{S},\mathcal{Y})$ be the regions of $(S,Y)$ of interest, and let $D_{i}=1(S_{i}>0)$. For example, $\mathcal{Y} = [q_0,q_1]$, where $q_0$ and $q_1$ are quantiles of $Y$. For
each $s\in\mathcal{S}$ and $y\in\mathcal{Y}$, we construct the indicators
$J_{i}^{s}=1(S_{i}\leq s)$, $\bar{J}_{i}^{s}=1-J_{i}^{s}$, $I_{i}^{y}=1(Y_{i}\leq y)$,
and $\bar{I}_{i}^{y}=1-I_{i}^{y}$. Note that $D_{i}=\bar{J}_{i}^{0}$.
We replace the arguments of $(\mu(s),\nu(y),\rho(s,y))$  by subscripts to lighten the
notation, that is, $(\mu_{s},\nu_{y},\rho_{sy})=(\mu(s),\nu(y),\rho(s,y))$. Estimation is based on forming the joint conditional likelihood of $J_{i}^{s}$ and $I_{i}^{y}$ given $Z_i$ taking into account that $I_{i}^{y}$ is only observed when  $D_{i}=1$. 
Thus, let $\xi_{sy} := (\mu_0,\mu_s,\nu_y,\rho_{0y},\rho_{sy})$, the likelihood is
    \begin{align*}
      f_i(\xi_{sy})=\Phi(-Z_i^\prime\mu_0)^{1-D_i}&\times \left[\Phi_{2}(Z_{i}^{\prime}\mu_{0},-X_{i}^{\prime}\nu_{y};-g(X_{i}^{\prime}\rho_{0y})) - \Phi_{2}(Z_{i}^{\prime}\mu_{s},-X_{i}^{\prime}\nu_{y};-g(Z_{i}^{\prime}\rho_{sy})) \right]^{D_{i}I^y_{i} J^s_{i}}
\\ 
      &\times\left[\Phi_{2}(Z_{i}^{\prime}\mu_{0},X_{i}^{\prime}\nu_{y};g(X_{i}^{\prime}\rho_{0y})) - \Phi_{2}(Z_{i}^{\prime}\mu_{s},X_{i}^{\prime}\nu_{y};g(Z_{i}^{\prime}\rho_{sy}))\right] ^{D_{i}\bar I^y_{i} J^s_{i}}
      \\ 
     &\times \Phi_{2}(Z_{i}^{\prime}\mu_{s},-X_{i}^{\prime}\nu_{y};-g(Z_{i}^{\prime}\rho_{sy}))^{D_{i} I^y_{i} \bar J^s_{i}}     \\
      & \times    \Phi_{2}(Z_{i}^{\prime}\mu_{s},X_{i}^{\prime}\nu_{y};g(Z_{i}^{\prime}\rho_{sy}))^{D_{i} \bar I^y_{i}\bar J^s_{i}}.        
      \end{align*}


\begin{algorithm}[Three-Step CDR Method]\label{alg:est}
    Let $\overline{\mS}$ and $\overline{\mY}$ be finite grids covering $\mS$ and $\mY$, respectively. \textbf{Step 1.} For each $s\in \{0\} \cup \overline{\mS}$, estimate $\mu_{s}$
by running probit regressions of $J_{i}^{s}$ on $Z_{i}$.
\begin{align*}
\hat{\mu}_{s} & =\argmax_{\mu_{s}}\dfrac{1}{n}\sum_{i=1}^{n}\bar{J}_{i}^{s}\log\Pr(S_{i}>s \mid Z_i)+J_{i}^{s}\log\Pr(S_{i}\leq s \mid Z_i)\\
 & =\argmax_{\mu_{s}}\dfrac{1}{n}\sum_{i=1}^{n}\bar{J}_{i}^{s}\log\Phi(Z_{i}^{\prime}\mu_{s})+J_{i}^{s}\log\Phi(-Z_{i}^{\prime}\mu_{s}).
\end{align*}
\textbf{Step 2.} For each $y \in \overline{\mY}$, estimate $\theta_{y}=(\nu_{y},\rho_{0y})$
by a  probit regression with sample selection correction.
\begin{align*}
\hat{\theta}_{y} & =\argmax_{\theta_{y}}\dfrac{1}{n}\sum_{i=1}^{n}D_{i}\left[\bar{I}_{i}^{y}\log P(S_{i}>0,Y_{i}>y \mid Z_i) +I_{i}^{y}\log P(S_{i}>0,Y_{i}\leq y \mid Z_i)\right]\\
 & =\argmax_{\theta_{y}}\dfrac{1}{n}\sum_{i=1}^{n}D_{i}\left[\bar{I}_{i}^{y}\log\Phi_{2}(Z_{i}^{\prime}\hat{\mu}_{0},X_{i}^{\prime}\nu_{y};g(X_{i}^{\prime}\rho_{0y}))+I_{i}^{y}\log\Phi_{2}(Z_{i}^{\prime}\hat{\mu}_{0},-X_{i}^{\prime}\nu_{y};-g(X_{i}^{\prime}\rho_{0y}))\right].
\end{align*}
\textbf{Step 3.} For each $(s,y) \in \overline{\mS} \times \overline{\mY}$, estimate
$\rho_{sy}$ by a bivariate probit regression,
\begin{align*}
\hat{\rho}_{sy}= & \argmax_{\rho_{sy}}\dfrac{1}{n}\sum_{i=1}^{n}D_{i}\Big[\bar{J}_{i}^{s}\bar{I}_{i}^{y}\log P(S_{i}>s,Y_{i}>y \mid Z_i)+\bar{J}_{i}^{s}I_{i}^{y}\log P(S_{i}>s,Y_{i}\leq y \mid Z_i)\\
 & \;\;\;\;\;\;\;\;\;\;\;\;\;\;\;\;\;+J_{i}^{s}\bar{I}_{i}^{y}\log P(0 < S_{i}\leq s,Y_{i}>y \mid Z_i)+J_{i}^{s}I_{i}^{y}\log P(0 < S_{i}\leq s,Y_{i}\leq y \mid Z_i)\Big]\\
= & \argmax_{\rho_{sy}}\dfrac{1}{n}\sum_{i=1}^{n}D_{i}\Big[\bar{J}_{i}^{s}\bar{I}_{i}^{y}\log\Phi_{2}(Z_{i}^{\prime}\hat{\mu}_{s},X_{i}^{\prime}\hat{\nu}_{y};g(Z_{i}^{\prime}\rho_{sy}))+\bar{J}_{i}^{s}I_{i}^{y}\log\Phi_{2}(Z_{i}^{\prime}\hat{\mu}_{s},-X_{i}^{\prime}\hat{\nu}_{y};-g(Z_{i}^{\prime}\rho_{sy}))\\
 & \;\;\;\;\;\;\;\;\;\;\;\;\;\;\;\;\;+J_{i}^{s}\bar{I}_{i}^{y}\log\left\{ \Phi_{2}(Z_{i}^{\prime}\hat{\mu}_{0},X_{i}^{\prime}\hat{\nu}_{y};g(X_{i}^{\prime}\hat{\rho}_{0y}))-\Phi_{2}(Z_{i}^{\prime}\hat{\mu}_{s},X_{i}^{\prime}\hat{\nu}_{y};g(Z_{i}^{\prime}\rho_{sy}))\right\} \\
 & \;\;\;\;\;\;\;\;\;\;\;\;\;\;\;\;\;+J_{i}^{s}I_{i}^{y}\log\left\{ \Phi_{2}(Z_{i}^{\prime}\hat{\mu}_{0},-X_{i}^{\prime}\hat{\nu}_{y};-g(X_{i}^{\prime}\hat{\rho}_{0y}))-\Phi_{2}(Z_{i}^{\prime}\hat{\mu}_{s},-X_{i}^{\prime}\hat{\nu}_{y};-g(Z_{i}^{\prime}\rho_{sy}))\right\} \Big].
\end{align*}
\end{algorithm}

The estimation algorithm is summarized in Algorithm \ref{alg:est}. We break down the maximization of the likelihood in three steps, which are illustrated in Figure \ref{fig1} in SM \ref{app:main-figure}. The first step is a probit regression  to estimate $\mu_{s}$ for each  $s \in \{0\} \cup\mS$. When $s=0$, we are estimating the probability
of employment, which is identical to the first step of the estimation algorithm
in CFL. The second step is a  probit regression with sample
selection correction to estimate $(\nu_{y},\rho_{0y})$ for each $y \in \mathcal{Y}$, which is also the same as the second step in CFL.
As mentioned previously, the likelihood function incorporates adjustment
to the sample selection. The final step is new relative to CFL and runs a bivariate probit regression
to estimate $\rho_{sy}$ for each $(s,y) \in \mS \times \mathcal{Y}$.
This step estimates the local correlation
parameters of the joint distribution plugging-in the parameters of the marginal distributions of $S^{*}$ and $Y^{*}$ estimated
in the first and second steps. See SM \ref{app:main-remark-remark1} for additional implementation details.

\begin{remark}[Comparison with \cite{chen2025distribution}]
The estimation procedures reflect the different identification strategies. Both have 3 steps. Step 1 is the same in both approaches, but Step 2 and 3 are different. Thus, their Step 2 pools multiple thresholds for $S$, whereas we only use one threshold. The source of this difference is that we impose the sorting exclusion restriction at $S=s_0$, whereas they impose it for every value of $S$. Step 3 also differs in the expression of the conditional log-likelihood. Ours has 4 components because we use all the uncensored observations, whereas theirs has 2 components because they only use uncensored observations with value of $S$ above the threshold, that is $\bar J_i^s =1$ in our notation.   \qed

\end{remark}

Estimators of the functionals of interest in Section \ref{sec:functional} are constructed
from the estimators of the parameters using the plug-in method. For
example, the marginal distributions of the latent outcome and selection
variable $(Y^{*},S^{*})$ are estimated with
\[
\hat{F}_{Y^{*}}(y)=\dfrac{1}{n}\sum_{i=1}^{n}\Phi(-X_{i}^{\prime}\hat{\nu}(y))\;\text{for }y\in\overline{\mY},\;\hat{F}_{S^{*}}(s)=\dfrac{1}{n}\sum_{i=1}^{n}\Phi(-Z_{i}^{\prime}\hat{\mu}(s))\;\text{for }s\in\overline{\mS}
\]
and the distribution of observed wage by worker with
type is
\[
\hat{F}_{Y}(y\mid\underline{s}<S^{*}\leq\overline{s})=\frac{\hat{F}_{S^{*},Y^{*}}(\overline{s},y)-\hat{F}_{S^{*},Y^{*}}(\underline{s},y)}{\hat{F}_{S^{*}}(\overline{s})-\hat{F}_{S^{*}}(\underline{s})},\;0\leq\underline{s}<\bar{s},y\in\overline{\mY}
\]
where $\hat{F}_{S^{*},Y^{*}}(s,y)=\dfrac{1}{n}\sum_{i=1}^{n}\Phi_{2}(-Z_{i}^{\prime}\hat{\mu}(s),-X_{i}^{\prime}\hat{\nu}(y);g(Z_{i}^{\prime}\hat{\rho}(s,y)))$.

\subsection{Uniform Inference}

We present asymptotic theory for the estimator of $(s,y) \mapsto \rho_{sy}$, the new function-valued parameter of interest, and show how to construct confidence bands
for this parameter using multiplier bootstrap. The theory in SM \ref{app:proof} applies to all parameters in the model, $(\eta_{sy},\rho_{sy})$, where $\eta_{sy}:=(\mu_0,\mu_{s},\nu_y,\rho_{0y})$. We only present the result for $\rho_{sy}$ in the main text as it is the most novel result relative to CFL.  

Let $L_{1}(\mu_s)$, $L_{2}(\theta_y,\mu_0)$, $L_{3}(\rho_{sy},\eta_{sy})$, $H_{1s}$, $H_{2y}$, $H_{3sy}$ and $\Sigma_{\rho}(s,y)$ be the functions defined in SM \ref{app:expressions}. We make the following assumption to derive the asymptotic distribution of $\hat \rho_{sy}$. 

\begin{assumption}[Model and Sampling]\label{ass:fclt} (1) Random sampling: $\{(S^*_i, Y^*_i,Z_i)\}_{i=1}^n$ is a sequence of independent and identically distributed copies of $(S^*,Y^*,Z)$. We observe $S = \max(S^*,0)$ and $Y = Y^*$ if $S > 0$.
(2) Model: the joint distribution of $(S^*,Y^*)$ conditional on $Z$ follows the BDR model \eqref{eq:drmodel} with the exclusion restrictions. (3) The support of $Z$, $\mathcal{Z}$,  is a  compact set.  (4) The sets $\mathcal{Y}$ and $\mathcal{S}$
are bounded intervals, which are compact subsets of the supports of $Y$ and $S$. 
The conditional density function $f_{S^* \mid Z}(s \mid z)$ exists, is uniformly bounded above, and is uniformly continuous in $(s,z)$ on $\mathcal{S} \times \mathcal{Z}_1$, where $\mathcal{Z}_1$ is the support of $Z$ conditional on $S>0$, and the conditional density function $f_{Y^* \mid X}(y \mid x)$ exists, is uniformly bounded above, and is uniformly continuous in $(y,z)$ on $\mathcal{Y}\times \mathcal{X}_1$, where $\mathcal{X}_1$ is the support of $X$ conditional on $S>0$.\footnote{See SM \ref{app:main-remark-remark2}  for discrete $Y$ and/or $S$ and a discussion on this condition.} (5) Identification and non-degeneracy: the equations $\Ep[\partial_{\mu_s} L_1(\tilde \mu_s)] = 0$, $\Ep[\partial_{\theta_y} L_2(\tilde \theta_y, \tilde \mu_0)] = 0$  and $\Ep[\partial_{\rho_{sy}} L_3(\tilde \rho_{sy}, \tilde \eta_{sy})] = 0$ posses a unique solution at $(\tilde \mu_s, \tilde \theta_y, \tilde \rho_{sy}) = (\mu_s, \theta_y,  \rho_{sy})$ that lies in the interior of a compact set $\mathcal{M} \times \Theta \times \mathcal{R} \subset \mathbb{R}^{d_{\mu} + d_{\theta} + d_{\rho}}$ for all $y \in \mathcal{Y}$ and $s \in \mathcal{S}$; and the minimum absolute eigenvalues of the matrices $H_{1s}$, $H_{2y}$, $H_{3sy}$ and $\Sigma_{\rho}(s,y)$ are bounded away from zero uniformly over $y \in \mathcal{Y}$ and $s \in \mathcal{S}$.
\end{assumption}


Let $\mathcal{SY}:= \mS \times \mY$, $\ell^{\infty}(\mathcal{SY})^{d_{\rho_{sy}}}$ be the set of bounded
functions on $\mathcal{SY}$, $\leadsto$ denote weak convergence (in distribution), and $\to_{\Pr}$ denote convergence in probability. The first main result of this section is a functional central limit theorem for $\hat{\rho}_{sy}$. 
\begin{thm}[FCLT for $\hat \rho_{sy}$]\label{thm:fclt} Under Assumption \ref{ass:fclt},
\[
\sqrt{n}(\hat{\rho}_{sy}-\rho_{sy})=-H_{3sy}^{-1}\sqrt{n}\left(S_{3sy}+J_{3sy}(\hat{\eta}_{sy}-\eta_{sy})\right)+o_{P}(1)\leadsto Z_{\rho_{sy}}  \text{ in } \ell^{\infty}(\mathcal{SY})^{d_{\rho_{sy}}},
\]
where $S_{3sy}  :=  \partial_{\rho_{sy}}L_{3}(\rho_{sy},\eta_{sy})$, 
$H_{3sy}  :=   \Ep \left[\partial_{\rho_{sy}\rho_{sy}^{\prime}}L_{3}(\rho_{sy},\eta_{sy})\right]$, 
$J_{3sy}  :=  \Ep \left[\partial_{\rho_{sy}\eta_{sy}^{\prime}}L_{3}(\rho_{sy},\eta_{sy})\right]$, and $(s,y)\mapsto Z_{\rho_{sy}}$ is a zero-mean Gaussian process with
uniformly continuous sample paths and covariance function 
$\Sigma_{\rho_{sy}\rho_{s'y'}} = \Ep[Z_{\rho_{sy}} Z_{\rho_{s'y'}}],$
for $(s,y),(s',y')\in\mathcal{SY}$. Also, $\hat{\Sigma}_{\rho}(s,y)$, the estimator of $\Sigma_{\rho}(s,y) := \Sigma_{\rho_{sy}\rho_{sy}}$ defined in \eqref{eq:se} is uniformly consistent, that is,
    $\sup_{(s,y) \in \mathcal{SY}} \|\hat{\Sigma}_{\rho}(s,y) - \Sigma_{\rho}(s,y) \| \to_{\Pr} 0$. The analytical expressions of $S_{3sy}$, $H_{3sy}$,  $J_{3sy}$ and $\Sigma_{\rho}(s,y)$ are given in SM \ref{app:expressions}.
\end{thm}
\medskip

We show how to construct confidence bands for function-valued
parameters that can be used to test functional hypotheses
such as the entire function being zero, non-negative or constant.
To explain the construction, consider the case where the functional
of interest is a linear combination of the model parameter $\rho_{sy}$
, that is the function $(s,y)\mapsto c^{\prime}\rho_{sy}$, where
$c\in\mathbb{R}^{d_{\rho_{sy}}}$ and $(s,y)\in\mathcal{SY}$.
The set $\textrm{CB}_{p}(c^{\prime}\rho_{sy})$ is a uniform (asymptotic)
$p$-confidence band for $c^{\prime}\rho_{sy}$ on $\mathcal{SY}_0 \subseteq \mathcal{SY}$ if $\Pr\left[c^{\prime}\rho_{sy}\in \textrm{CB}_{p}(c^{\prime}\rho_{sy}),\forall(s,y)\in\mathcal{SY}_0\right]\to p \text{ as } n \to \infty$.

Let  $\hat{\Sigma}_{\rho}(s,y)$ be a uniformly consistent estimator of $\Sigma_{\rho}(s,y)$, such as the one given in \eqref{eq:se}.
We construct $\textrm{CB}_{p}(c^{\prime}\rho_{sy})=c^{\prime}\hat{\rho}_{sy}\pm cv(p)\sqrt{c'\hat{\Sigma}_{\rho}(s,y)c/n}$,
where $cv(p)$ is a consistent estimator of the $p$-quantile of the
maximal t-statistic $t_{\mathcal{SY}_0}=\sup_{(s,y)\in\mathcal{SY}_0} \sqrt{n}\frac{|c^{\prime}\hat{\rho}_{sy}-c^{\prime}\rho_{sy}|}{\sqrt{c'\hat{\Sigma}_{\rho}(s,y)c}}$. We can obtain the critical value from the limit distribution of the stochastic process in Theorem \ref{thm:fclt}. 

In practice, it is convenient
to estimate the critical value using resampling methods. The multiplier
bootstrap method of \citet{gine1984some} is computationally attractive in
our setting because it does not require parameter re-estimation and
therefore avoids the nonlinear optimization in our estimation algorithm.
The multiplier bootstrap is implemented using the following algorithm.
Confidence bands for other functionals of the model parameter can
be constructed using a similar  method.

\medskip

\begin{algorithm}[Multiplier Bootstrap]\label{alg:mboot}
    For $b\in1,\ldots,B$ and a finite grid $\overline{\mathcal{SY}}_0$ covering $\mathcal{SY}_0$, 
\textbf{Step 1.} Draw bootstrap multipliers $\{\omega_{i}^{b}:1\leq i\leq n\}$
independently from the data and normalized them to have zero mean,
$\omega_{i}^{b}=\tilde{\omega}_{i}^{b}-\sum_{i=1}^{n}\tilde{\omega}_{i}^{b}/n,\ \ \tilde{\omega}_{i}^{b}\sim\text{ i.i.d. }\mathcal{N}(0,1)$.\\
\textbf{Step 2.} Obtain a bootstrap draw of the estimator of $\rho_{sy}$ as
$\hat{\rho}_{sy}^{b}=\hat{\rho}_{sy}+n^{-1}\sum_{i=1}^{n}\omega_{i}^{b} \hat \psi_{3sy,\hat \xi}(W_i)$, where $\hat \psi_{3sy,\hat \xi}(W_i)$ is the estimator of the influence function of $\hat{\rho}_{sy}$ given in \eqref{eq:influence}.\\
\textbf{Step 3.} Construct bootstrap realization of maximal t-statistic $t_{\overline{\mathcal{SY}}_0}^{b}=\max_{(s,y)\in\overline{\mathcal{SY}}_0} \sqrt{n} \frac{|c'\hat{\rho}_{sy}^{b}-c'\widehat{\rho}_{sy}|}{\sqrt{c'\hat{\Sigma}_{\rho}(s,y)c}}$, 
where $\hat{\Sigma}_{\rho}(s,y)$ be the estimator of $\Sigma_{\rho}(s,y)$ given in \eqref{eq:se}.
Compute the critical value $cv(p)$ as the simulation $p$-quantile
of $t_{\overline{\mathcal{SY}}_0}^{b}$, that is, $cv(p)=p-\text{quantile of }\left\{t_{\overline{\mathcal{SY}}_0}^{b}:1\leq b\leq B\right\}$.
\end{algorithm}

We make the following assumption about the multipliers:
\begin{assumption}[Multiplier Bootstrap]\label{ass:mb} The multipliers  $(\omega_{1},...,\omega_{n})$ are i.i.d. draws from a
random variable $\omega \sim \mathcal{N}(0,1)$, and are
independent of  $\{(S^*_i, Y^*_i,Z_i)\}_{i=1}^n$ for all $n$.
\end{assumption}

We establish a functional central limit theorem for the bootstrap for  $\hat \rho_{sy}$. 
  Here we use $\leadsto_{\Pr}$ to denote
bootstrap consistency, i.e. weak convergence conditional on the data in
probability, which is formally defined in \cite{van1996weak}.\footnote{We recall this definition in SM \ref{app:notation}.}

\begin{thm}[Multiplier Bootstrap FCLT for $\hat \rho_{sy}$]\label{thm:fclt_boot}
Under Assumptions \ref{ass:fclt} and \ref{ass:mb}, $\sqrt{n}(\hat{\rho}^b_{sy}-\hat \rho_{sy}) \leadsto_{\Pr} Z_{\rho_{sy}} \text{ in } \ell^{\infty}(\mathcal{SY})^{d_{\rho_{sy}}}$, where $Z_{\rho_{sy}}$ is the same Gaussian process as in Theorem \ref{thm:fclt}. In particular, $\Pr\left[c^{\prime}\rho_{sy}\in \textrm{CB}_{p}(c^{\prime}\rho_{sy}),\forall(s,y)\in\overline{\mathcal{SY}}_0\right]\to p \text{ as } n \to \infty,
$ where $\textrm{CB}_{p}(c^{\prime}\rho_{sy})$ is the confidence band obtained in Algorithm \ref{alg:mboot}.
\end{thm}
In SM \ref{app:simulation}, we report results from a Monte Carlo simulation calibrated to the empirical application that examines the finite sample performance of our estimators and multiplier bootstrap confidence bands.

\section{Employment Sorting and Wage Decompositions in the U.K.}\label{sec:empirics}
We analyze 
sorting into full-time and overtime work across gender, marital
status, and time in the UK labor market. Furthermore, we conduct wage
decompositions to compare the wage distributions of men and women
for full-time and overtime works. 

\subsection{Data}

We use the same dataset as CFL, which contains repeated cross-sectional
observations spanning from 1978 to 2013 \citep{data}. The sources of the data are the
Family Expenditure Survey (FES) from 1978 to 2001, the Expenditure
and Food Survey (EFS) from 2002 to 2007, and the Living Costs and
Food Survey (LCFS) from 2008 to 2013.  The sample
is constructed similarly to the previous studies that used the FES,
such as \citet{gosling2000changing}, \citet{blundell2003interpreting},
and \citet{blundell2007changes}. It includes individuals aged 23 to
59 years, excluding full-time students, self-employed individuals,
those married with an absent spouse, and those with missing education
or employees with missing wage information. This results in a sample
of 258,900 observations, with 139,504 women and 119,396 men.

The outcome variable of interest, $Y$, is the logarithm of the real
hourly wage rate. This is calculated as the ratio of weekly usual
gross nominal earnings to weekly usual work hours, adjusted for
inflation using the UK quarterly retail price index. The selection
variable, $S$, is the weekly usual work hours. The covariates,
$X$, include indicators for age at which individuals left school,
a quartic polynomial in age, an indicator for being married or cohabiting,
the number of children categorized by age, survey year indicators,
and region. Table \ref{table1} in SM \ref{app:B1} presents descriptive statistics for these variables.

The instrumental variable, $Z_{1}$, is the potential out-of-work
income benefit interacted with the marital status indicator, as previously
used in \citet{blundell2003interpreting}, \citet{blundell2007changes} and CFL.
This variable is constructed using
the Institute of Fiscal Studies (IFS) tax and welfare-benefit simulation
model (TAXBEN) for each individual. TAXBEN calculates the income of a tax unit if the
individual were out-of-work, factoring in eligible unemployment and
housing benefits. The primary sources of variation are the household
demographic composition, housing costs, and policy changes across
regions over time. We assume the outcome and selection sorting exclusion restrictions at $s_0=0$, that is, we assume that, conditional on observed covariates, the offered wage and the
dependence between offered wage and desired work hours at $s=0$ do not depend
on the benefit level. While the plausibility of the exclusion restriction
is controversial, we assume these conditions are satisfied and refer
to \citet{blundell2003interpreting} and \citet{blundell2007changes}
for a discussion on the validity of outcome exclusion restriction.
For the selection sorting exclusion restriction, we motivate its use
by noting the widespread application of the classical HSM in empirical
research, which trivially satisfies the restriction.

\subsection{Empirical specifications}

We estimate the censored distribution regression model separately
for men and women. The selection (work hours) equation includes all the
covariates mentioned earlier, while the outcome (wage) equation is identical
except for the excluded covariates. Given that estimating the parameter
of the selection sorting function is notoriously more challenging
than estimating the parameters of other equations, we consider four simplified specifications for the covariate vector \(z\) in the index \(z'\rho(s,y)\):
\[
z=
\begin{cases}
1,                    & \text{Specification 1},\\
(m,\,1-m)',           & \text{Specification 2},\\
(1,\,t)',             & \text{Specification 3},\\
(1,\,t,\,m,\,tm)',    & \text{Specification 4},
\end{cases}
\qquad
m=\mathbf{1}\{\text{married}\},\quad
t=\text{year}-1978.
\]

Using these specifications, we estimate the selection sorting parameters $\rho(s,y)$ on $\mathcal{SY}_0 = \mathcal{S}_0  \times \mathcal{Y}_0 $
and plug-in these estimates to the selection sorting function $(s,y)\to g\left(z^{\prime}\rho(s,y)\right)$ for the population group of interest defined by the value of $z$. $\mathcal{S}_0 = \{0,34,40\}$, the thresholds corresponding to part-time, full-time and overtime work, and $\mathcal{Y}_0$ is the interval between the sample quantiles of $Y$ with orders $.1$ and $.9$. For example, using specification
2, we estimate the selection sorting function for married
men  $(s,y)\to g\left(\rho_{M}(s,y)\right)$ using the sample of
men. In Algorithms  1 and 2, we use a grid $\overline{\mathcal{SY}}_0 = \mathcal{S}_0 \times \overline{\mathcal{Y}}_0$, where $\overline{\mathcal{Y}}_0$ is a grid of sample quantiles of $Y$ with
indices $\left\{ 0.1,0.11,...,0.9\right\} $. We present 95\% confidence bands for the selection sorting
function using Algorithm 2 with $B=500$ bootstrap repetitions. 

\bigskip
\begin{figure} [ht!]
\centering     
\begin{subfigure}{0.49\textwidth}         
\includegraphics[width=\textwidth]{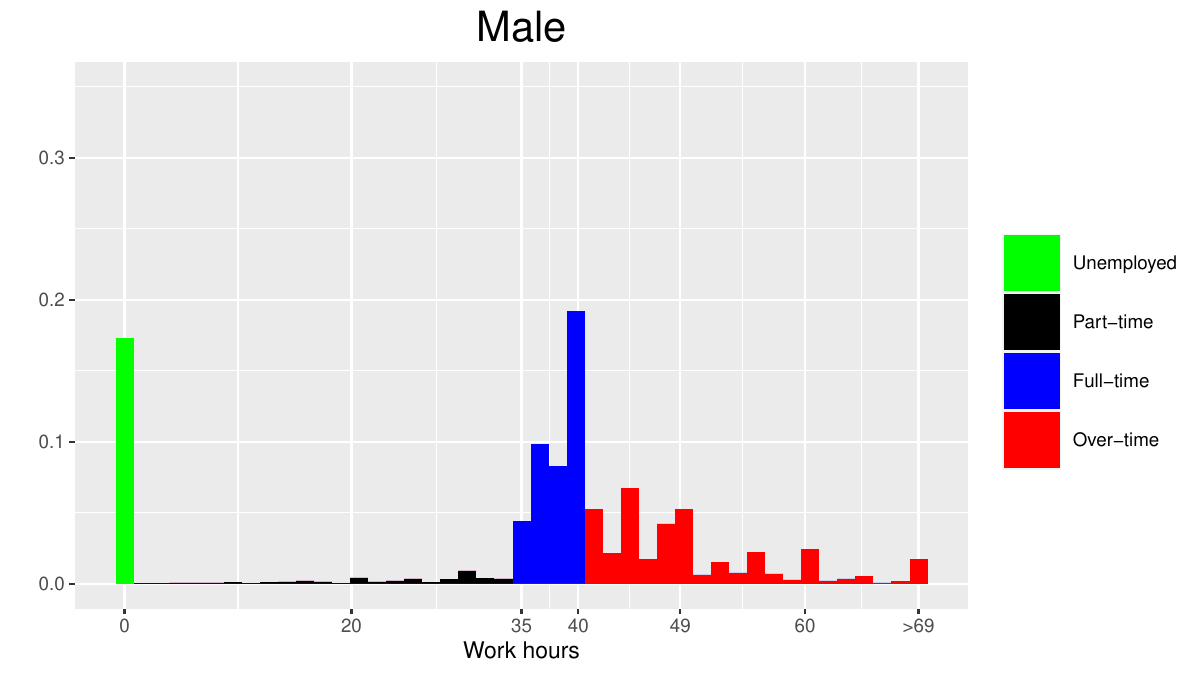}         
\end{subfigure}     
\begin{subfigure}{0.49\textwidth}         
\includegraphics[width=\textwidth]{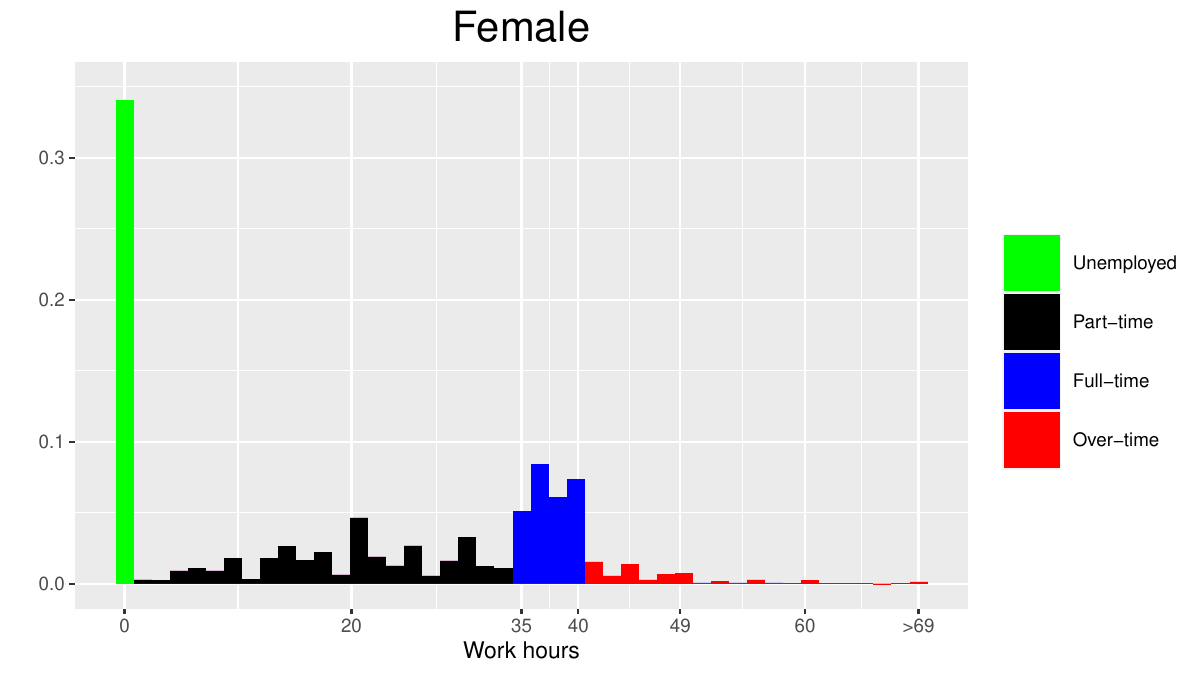}         
\end{subfigure}     
\caption{Histograms of work hours by gender} \label{fig2}
\end{figure}

Our specification of $\mathcal{S}_0$ involves the definition of part-time,
full-time, and overtime work, with thresholds set at 34 and
40 hours. This means we classify individuals working 34 hours or less
as part-time workers, those working from 35 to 40 hours as full-time
workers, and those working 41 hours or more as overtime workers. Although
we have not been able to find formal definitions of these worker types, we have adopted
thresholds that are commonly accepted in the literature (\citet{mulligan2008selection}, \citet{maasoumi2019gender}). Moreover, as shown in Figure \ref{fig2}, which presents histograms of work hours by gender, our data exhibit clear bunching at 34 and 40 hours. Table \ref{table1} of SM \ref{app:B1} shows significant gender differences in employment and work intensity. Employment rates are 83 percent for men and 66 percent for women. Among workers, 5, 50, and 45 percent of men work part-time, full-time, and overtime, respectively, compared with 50, 41, and 9 percent of women. These patterns motivate an analysis of gender-specific selection patterns across work hours.

\subsection{Main results}

\subsubsection{Employment sorting}

\begin{figure} [ht!]
\centering     
\includegraphics[width=\textwidth]{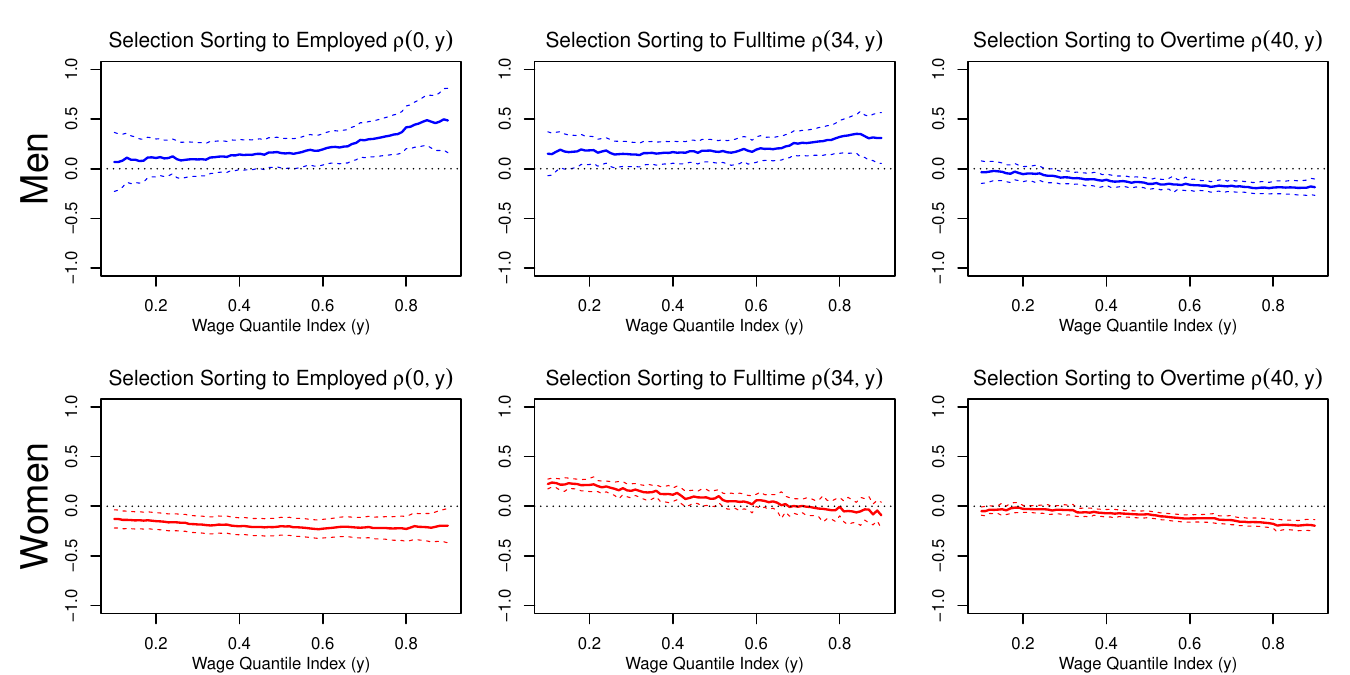}         
\caption{Selection sorting function using Specification 1} \label{fig3}
\end{figure}

Figure \ref{fig3} presents estimates of selection sorting functions using Specification 1. The results for selection sorting into employment in the first column replicate those reported in CFL. The estimated sorting functions indicate that selection into full-time and overtime work exhibit distinct patterns compared to selection into employment. Specifically, men show positive selection into employment and full-time work but negative selection into overtime work. In contrast, women exhibit negative selection into employment and overtime work but positive selection into full-time work. These findings showcase  how the censored selection model is able to uncover heterogeneous selection sorting patterns based on work hours, which cannot be captured using a binary selection model. In the subsequent section, we discuss  results using Specifications 2 to 4 to identify the primary drivers of the different sorting patterns observed in Specification 1, focusing on marital status and changes over time.

\begin{figure} [ht!]
\centering
\includegraphics[width=0.75\textwidth]{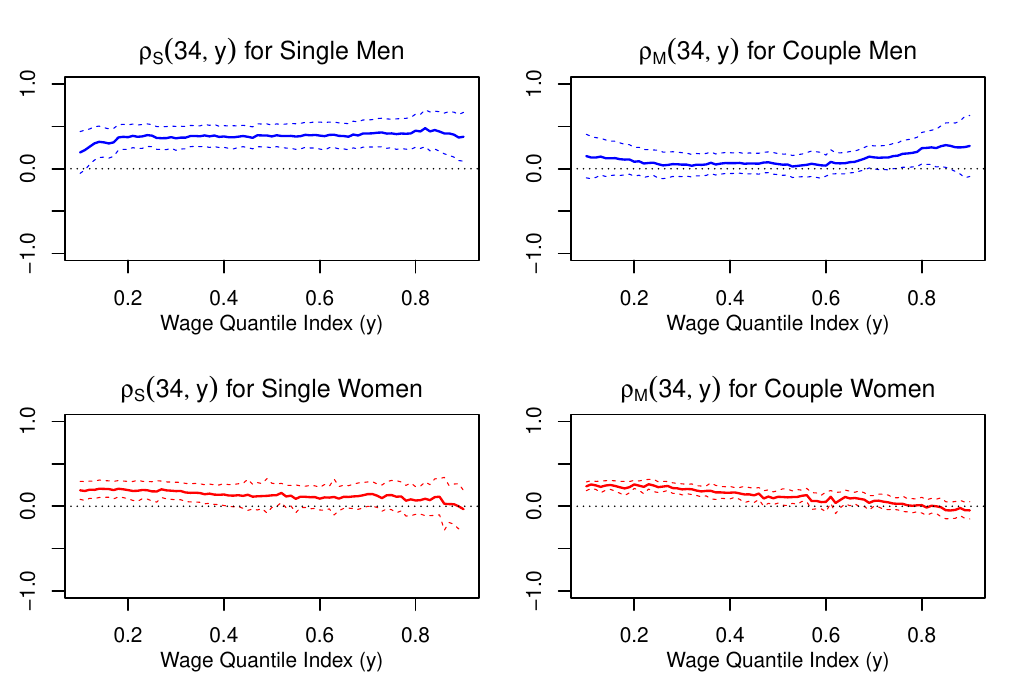}        
\caption{Selection sorting into full-time using Specification 2} \label{fig4}
\end{figure}

\begin{figure} [ht!]
\centering     
\includegraphics[width=.75\textwidth]{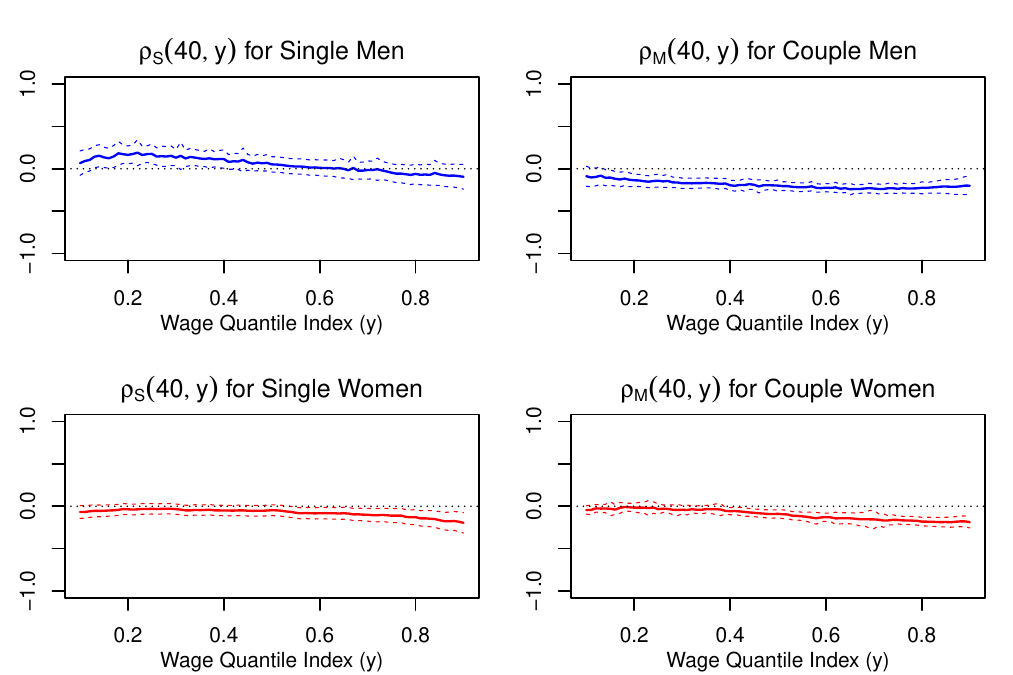}         
\caption{Selection sorting into overtime using Specification 2}  \label{fig5}
\end{figure}

Figures \ref{fig4} and \ref{fig5} present the results for selection into full-time and overtime work using Specification 2. We first discuss the findings for selection into full-time work. Figure \ref{fig3} shows that men exhibit positive sorting across the wage distribution, whereas women show positive sorting only at the lower wage quantiles. Figure \ref{fig4} indicates that these sorting patterns are particularly pronounced among single men and married women. We interpret this as a
result of single men and married women having greater flexibility
in selecting their work hours. Traditionally, men have been the primary
financial supporters of their families after marriage, which may allow
single men and married women more freedom in choosing work hours,
leading to significant positive sorting patterns. Thus, single men and married women are relatively more likely to work full time when they have relative comparative advantage based on unobservable characteristics, say ability.

To explain why married women exhibit positive selection only at the
bottom of the distribution, we adopt the concepts of voluntary and
involuntary part-time work. According to \citet{dunn2018chooses},
part-time workers can be classified into two categories based on their
reasons for working part-time. Involuntary part-time workers seek
full-time work but work part-time because they can only find
part-time jobs. In contrast, voluntary part-time workers prefer part-time
work due to non-economic reasons, such as childcare, school,
or medical issues. \citet{cam2012involuntary}, using UK Labour Force
Survey data, examined the determinants of involuntary part-time work.
Their findings suggest that women working part-time with dependent
children are more likely to be voluntary, especially when they are
in a couple. Based on this, we posit that single male part-time workers
are predominantly involuntary, meaning only those with high potential
wages are likely to work full-time. On the other hand, married women
in part-time work are a mix of voluntary and involuntary workers,
which explains why we observe a selection effect only at the bottom
of the distribution.

For selection into overtime work, Figure \ref{fig3} shows that both genders
exhibit a negative selection effect, which is more pronounced
at the top of the wage distribution. Figure \ref{fig5} further reveals that this
negative selection effect is more pronounced for married men and women. One
potential explanation for this pattern is that higher-wage jobs are more likely to be task-based in the sense that the work hours are determined by the time needed to complete the assigned tasks. Consequently,
workers who require overtime at high wage levels are more likely to be those
with relatively lower ability, which is reflected in the negative
selection effect observed in our estimates.

Figure \ref{fig6} in SM \ref{app:sorting} shows the selection sorting function using Specification 3, revealing that sorting
has become more positive and homogeneous across wage quantiles over
time. For women, this aligns with \citet{maasoumi2019gender}, who
showed that selection into full-time shifts from negative to positive
over time. However, while they estimate a single selection parameter
for each year, we estimate a selection sorting function across wage
quantiles. Figures \ref{fig10} and \ref{fig11} in the SM \ref{app:sorting}  indicate that these sorting
trends are primarily driven by single men and married women, similar
to the findings in Specifications 1 and 2. 


\subsubsection{Wage decomposition}

We next use the DR model to decompose changes in the distribution
of the observed wages between women and men for full-time and overtime
workers. The goal is to investigate the factors contributing to the
gender wage gap according to worker type. We do not include the result
for part-time workers, as there are too few part-time male workers
in our dataset, leading to unreliable results. 

We identify four components corresponding to different inputs of the
DR model: (1) wage structure $\nu(y)$; (2) selection sorting $\rho(s,y)$;
(3) selection structure $\mu(s)$; and (4) composition $F_{Z}$. Let
$Q_{Y\langle t,j,r,k\rangle}^{(\underline{s},\overline{s}]}$ represent
the (counterfactual) quantiles of workers with work hours within the
$(\underline{s},\overline{s}]$ interval, with $t$-wage structure,
$j$-hours sorting, $r$-hours structure, and $k$-composition. The
actual quantile in group $t$ corresponds to $Q_{Y\langle t,t,t,t\rangle}^{(\underline{s},\overline{s}]}$.
We consider two groups, labeled 0 and 1, representing different demographic
populations, such as women and men. The observed wage quantile difference
between group 1 and 0 can be decomposed as:
\begin{align*}
\underset{\text{Total}}{\underbrace{Q_{Y\langle1,1,1,1\rangle}^{(\underline{s},\overline{s}]}-Q_{Y\langle0,0,0,0\rangle}^{(\underline{s},\overline{s}]}}} & =\underset{\text{Wage Structure}}{\underbrace{[Q_{Y\langle1,1,1,1\rangle}^{(\underline{s},\overline{s}]}-Q_{Y\langle0,1,1,1\rangle}^{(\underline{s},\overline{s}]}]}}+\underset{\text{Selection Sorting}}{\underbrace{[Q_{Y\langle0,1,1,1\rangle}^{(\underline{s},\overline{s}]}-Q_{Y\langle0,0,1,1\rangle}^{(\underline{s},\overline{s}]}]}}\\
 & +\underset{\text{Selection Structure}}{\underbrace{[Q_{Y\langle0,0,1,1\rangle}^{(\underline{s},\overline{s}]}-Q_{Y\langle0,0,0,1\rangle}^{(\underline{s},\overline{s}]}]}}+\underset{\text{Composition}}{\underbrace{[Q_{Y\langle0,0,0,1\rangle}^{(\underline{s},\overline{s}]}-Q_{Y\langle0,0,0,0\rangle}^{(\underline{s},\overline{s}]}]}}.
\end{align*}
In this decomposition, the first term on the right-hand side captures
the wage structure effect, the second reflects the hours-wage
sorting effect, the third represents the hours structure
effect, and the fourth indicates the composition effect. This generalizes
the classical Oaxaca-Blinder decomposition (\citet{oaxaca1973male}, \citet{blinder1973wage}) to analyze distributional
gap of the outcome and to take sample selection into account. Furthermore,
since we are adopting censored selection rule, we can decompose wage
distribution gaps according to worker type by work hours, in contrast
to CFL decomposing wage distribution gap only of employed. While it is
well-known that the order of component extraction in such decompositions
can influence the results, we estimate the decomposition using different
orderings of the components and find that the main conclusions remain
consistent.

Let $F_{Y\langle t,j,r,k\rangle}^{(\underline{s},\overline{s}]}$
be the (counterfactual) distribution of wages, which can be expressed
as
\begin{footnotesize}
\[
F_{Y\langle t,j,r,k\rangle}^{(\underline{s},\overline{s}]}(y)=\frac{\int[\Phi_{2}(-z^{\prime}\mu_{r}(\overline{s}),-x^{\prime}\nu_{t}(y);g(z^{\prime}\rho_{j}(\overline{s},y)))-\Phi_{2}(-z^{\prime}\mu_{r}(\underline{s}),-x^{\prime}\nu_{t}(y);g(z^{\prime}\rho_{j}(\underline{s},y)))]dF_{Z_{k}}(z)}{\int[\Phi(-z^{\prime}\mu_{r}(\overline{s}))-\Phi(-z^{\prime}\mu_{r}(\underline{s}))]dF_{Z_{k}}(z)},
\]
\end{footnotesize}where $\nu_{t}$ is the coefficient of the wage equation in group
$t$, $\rho_{j}$ is the coefficient of the sorting function in group
$j$, $\mu_{r}$ is the coefficient of the selection equation in group
$r$, and $F_{Z_{k}}$ is the distribution of characteristics in group
$k$. We construct a plug-in estimator of $F_{Y\langle t,j,r,k\rangle}^{(\underline{s},\overline{s}]}(y)$
by appropriately using the estimates of model parameters and covariate
distributions from the two groups. Lastly, we use the generalized quantile operator
$Q_{\tau}(F)= \int_{0}^\infty 1(F(y) < \tau) \mathrm{d} y -  \int_{-\infty}^{0} 1(F(y) \geq \tau) \mathrm{d} y$,
to estimate $Q_{Y\langle t,j,r,k\rangle}^{(\underline{s},\overline{s}]}(\tau)$.\footnote{The operator $Q_\tau$ monotonically arrange the estimator of the distribution before applying the (left) inverse.} 

\begin{figure} [ht!]
\centering     
\includegraphics[width=.85\textwidth, height=7cm]{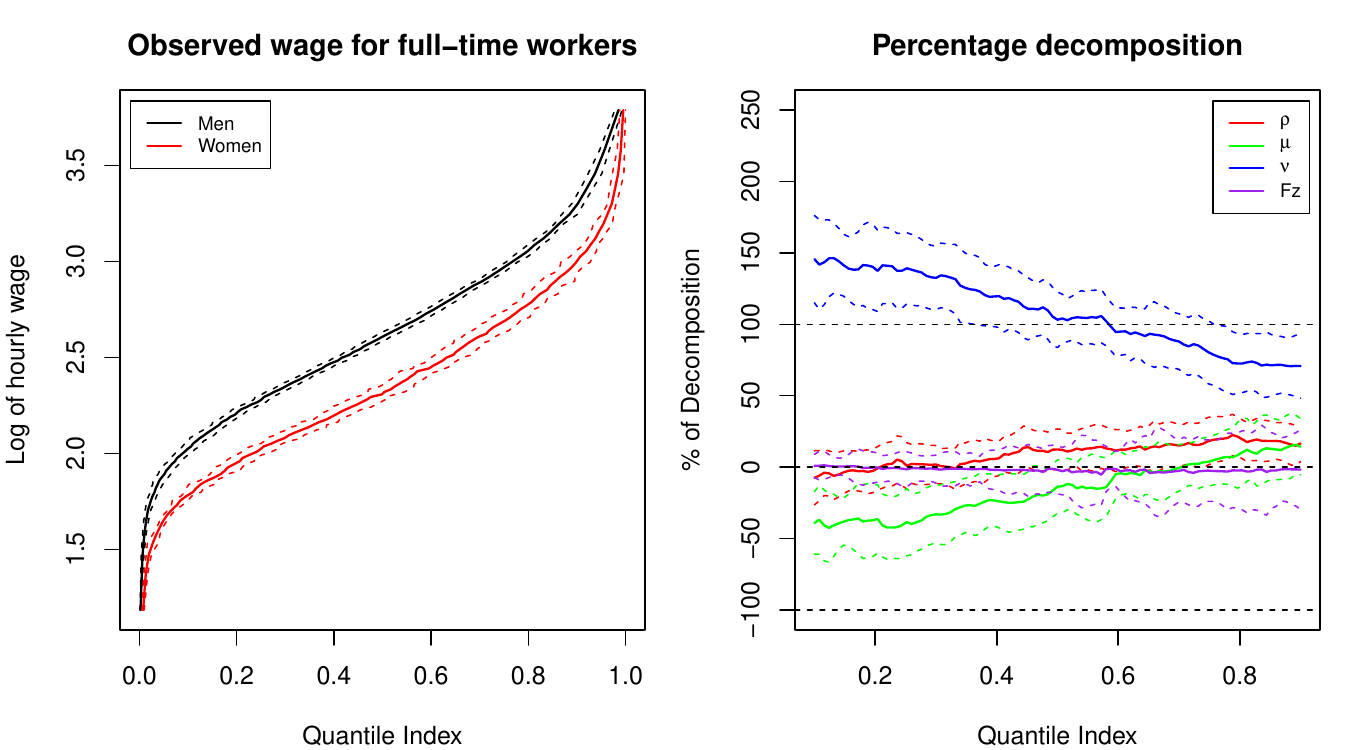}         
\caption{Quantile estimates of observed wages and decomposition between full-time working men and women in Specification 2} \label{fig7}
\end{figure}

\begin{figure} [ht!]
\centering     
\includegraphics[width=.85\textwidth, height=7cm]{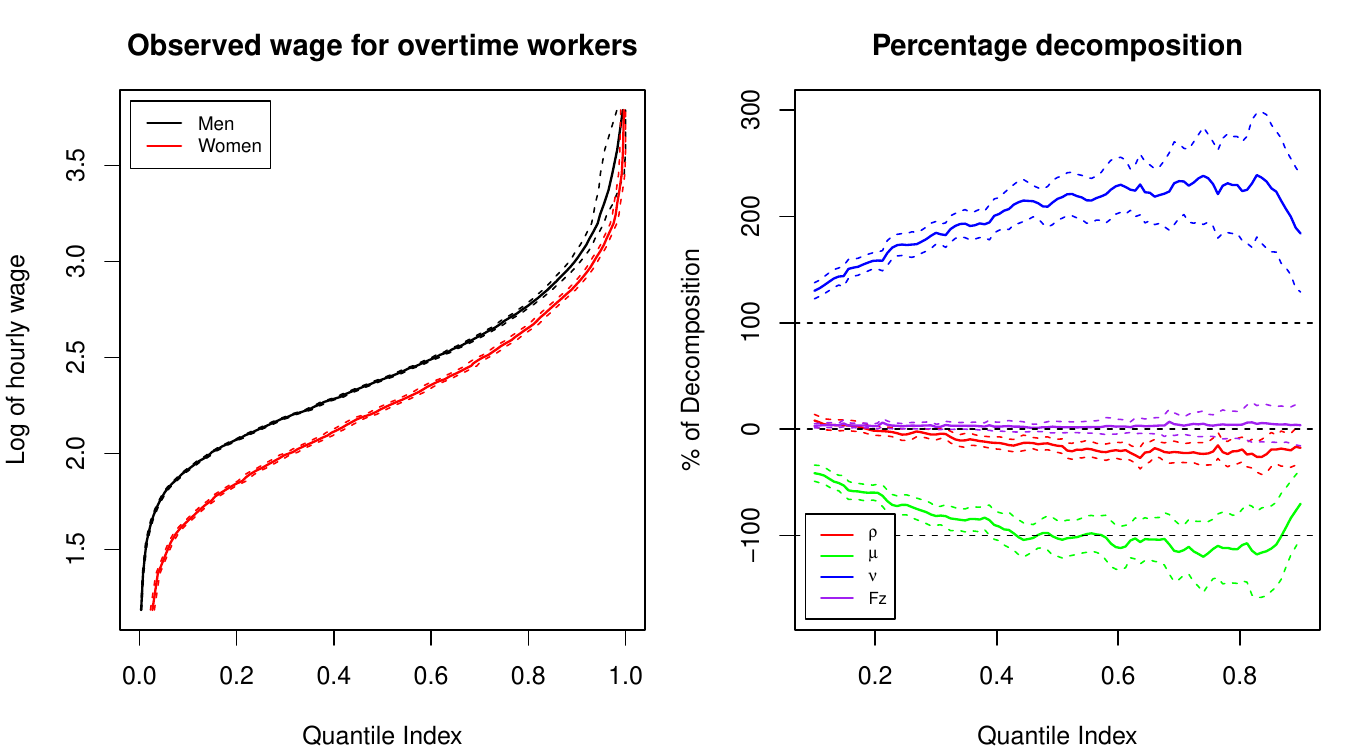}         
\caption{Quantile estimates of observed wages and decomposition between overtime working men and women in Specification 2} \label{fig8}
\end{figure}

Figures \ref{fig7} and \ref{fig8} present the estimated quantile functions for observed
wages among full-time and overtime working men and women, along with
the relative contributions of each component to the wage distribution
gap between men (group 1) and women (group 0) based on Specification
2. For both full-time and overtime workers, men exhibit higher wage quantiles than women. However, note that the wage gap at each quantile index is smaller for overtime workers compared to full-time workers starting from $\tau=0.14$ and onward. Specifically, the wage gap for full-time workers is up to 2.5 times larger than that for overtime workers at the same quantile index. In the decomposition results, a positive contribution indicates that the component accounts for a proportion of the estimated gap, while a negative contribution suggests that the component is associated with narrowing the estimated gap.

For full-time workers, most of the gender wage gap is attributed to
differences in the wage structure, that is, differences in returns
to observed charactersitics. However, the hours structure and
hours-wage sorting effects also explain an important portion of the gap.
At lower wage quantiles, the hours structure has a negative contribution.
This implies that women with similar observed characteristics to men
are less likely to work full-time, leading to a reduced observed wage
gap. At higher wage quantiles, the hours-wage sorting effect partially
explains the gap, similar to the decomposition result for the employed
in CFL.

For overtime workers, the percentages exceed $100$ or fall below
$-100$ because the wage distribution gaps generated by the counterfactual
distributions are larger than the observed gap, which is used as the
denominator. The decomposition procedure shows that if we impose the
women's wage structure onto the men's wage distribution, the resulting
counterfactual distribution yields much lower quantiles than the observed
quantiles for women.\footnote{Note that $[Q_{Y\langle1,1,1,1\rangle}^{(\underline{s},\overline{s}]}-Q_{Y\langle0,1,1,1\rangle}^{(\underline{s},\overline{s}]}]/[Q_{Y\langle1,1,1,1\rangle}^{(\underline{s},\overline{s}]}-Q_{Y\langle0,0,0,0\rangle}^{(\underline{s},\overline{s}]}] > 1$ if and only if $Q_{Y\langle0,1,1,1\rangle}^{(\underline{s},\overline{s}]} < Q_{Y\langle0,0,0,0\rangle}^{(\underline{s},\overline{s}]}$.} The largely negative aggregate selection effect
--- which includes both the hours structure and hours-wage sorting effects
--- suggests that selection behavior is a key factor in reducing
the observed wage gap. For example, men with similar observed characteristics
to women are more likely to work overtime, which lowers the observed
wage quantiles for men relative to women as both gender exhibit negative sorting
into overtime work. In fact, the estimated probability of working
overtime with male hours structure and male composition is 0.37, whereas
the counterfactual probability with female hours structure and male
composition is 0.07. This supports the idea that women with the same
observed characteristics as men are less likely to work overtime. The work hours decomposition result in SM \ref{app:work hours} shows that the majority of the work hours difference is explained by the hours structure rather than composition effects. This aligns with the wage decomposition result in that composition effects have little explanatory power in the gender wage gap.

\section{Conclusion}\label{sec:conclusion}

We propose a distribution regression model with censored
sample selection rule. 
Compared to the distribution regression model using a binary
selection rule, our approach allows for the examination of selection
effects based on both the outcome and selection variables. We
apply the developed model to wage and work hours  data from the U.K.
to investigate selection into full-time and overtime work. Additionally,
we conduct a decomposition of wage and work hours distributions by gender for these
worker types.

\subsection*{Disclosure statement}
The authors used ChatGPT for language editing and R code organization. The authors reviewed and verified the output and take full responsibility for the content of the published article.

\subsection*{Data Availability Statement}
The data that support the findings of this study are openly available upon registration in UKDS at https://doi.org/10.5255/UKDA-SN-9355-1, reference number SN 9355.


\bibliographystyle{econometrica}
\bibliography{Reference}

\newpage

\appendix
\begin{center}
{\large\bf SUPPLEMENTARY MATERIAL}
\end{center}
\section{Supplements to Section \ref{sec:dr}}

\subsection{Figure with Estimation Steps}\label{app:main-figure}

\begin{figure} [ht!]
\centering     
\begin{subfigure}{0.3\textwidth}         
\includegraphics[width=\textwidth]{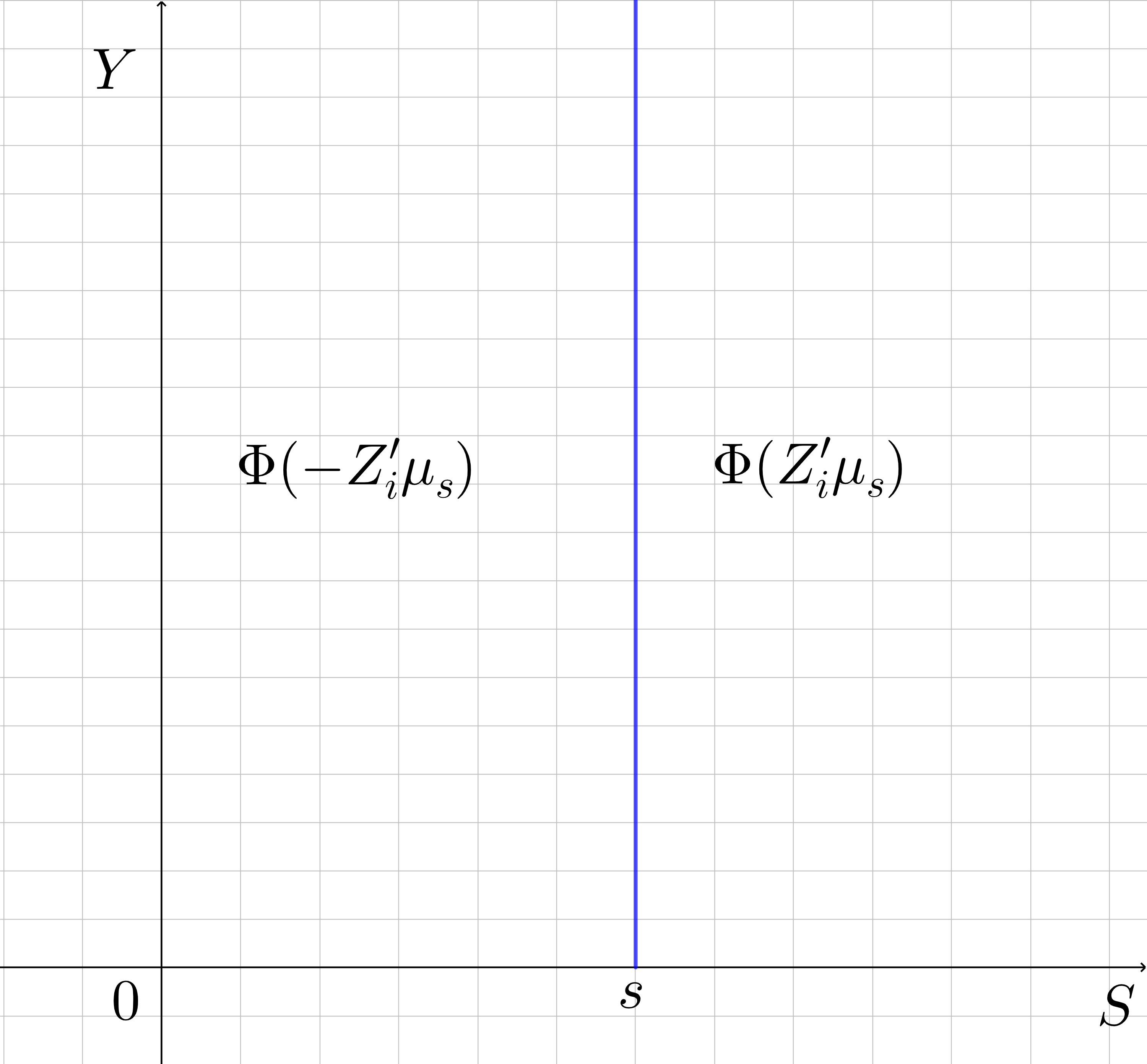}         
\caption{Step 1}     
\end{subfigure}     
\hfill     
\begin{subfigure}{0.3\textwidth}         
\includegraphics[width=\textwidth]{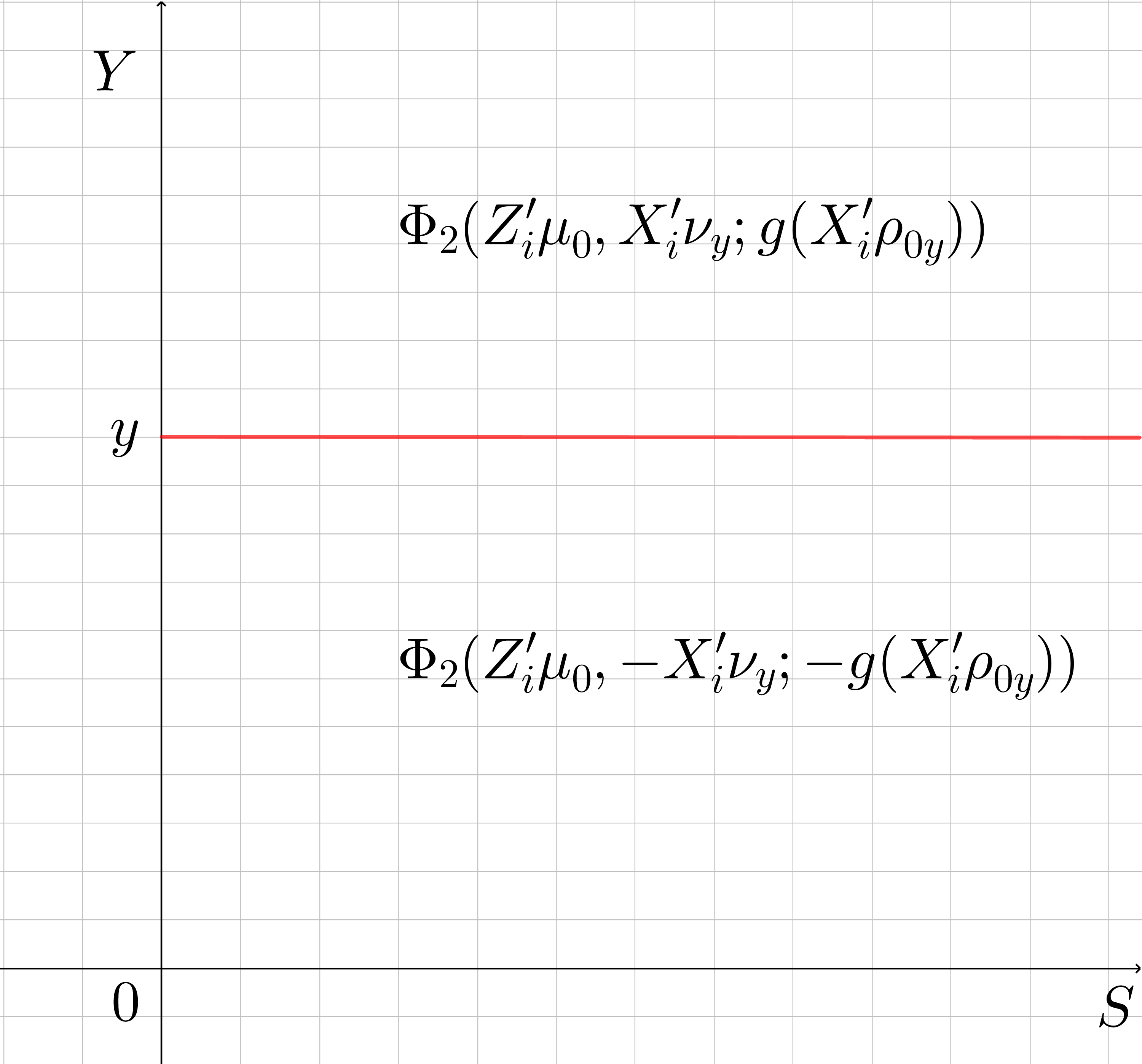}         
\caption{Step 2}     
\end{subfigure}     
\hfill     
\begin{subfigure}{0.3\textwidth}         
\includegraphics[width=\textwidth]{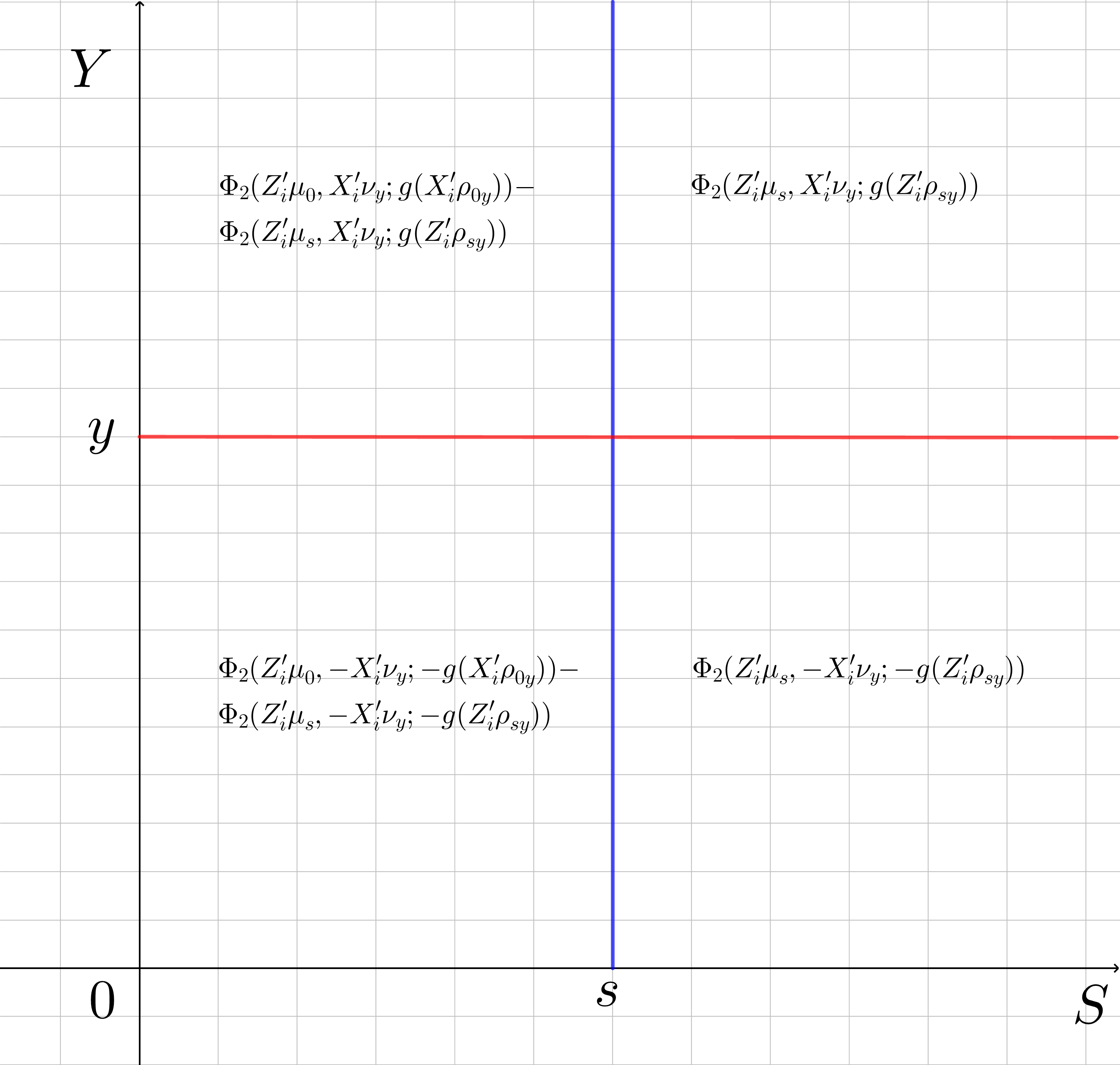}         
\caption{Step 3}     
\end{subfigure} 
\caption{Estimation steps} \label{fig1}
\end{figure}

\subsection{Computation}\label{app:main-remark-remark1}

When implementing Step 3 of Algorithm 1, we encountered the issue that some
probabilities were evaluated as negative in certain regions of the
parameter space. Specifically, the model predictions of $P(0 < S_{i}\leq s,Y_{i}>y  \mid Z_i)$ or $P(0 < S_{i}\leq s,Y_{i}\leq y \mid Z_i)$
were negative when the prediction of $P(S_{i} > 0,Y_{i}>y \mid Z_i)$ was smaller
than the prediction of $P(S_{i}> s,Y_{i}>y \mid Z_i)$ or the prediction of $P(S_{i} > 0,Y_{i}\leq y \mid Z_i)$ was
smaller than the prediction of $P(S_{i}> s,Y_{i}\leq y \mid Z_i)$, respectively. Using simulations, we confirmed
that this region of negative predicted probabilities occurs when the parameters are far from their true
values, suggesting it is unlikely to affect estimation.
However, to avoid computational problems, we adopted a smooth transformation
 to keep the estimated probabilities non-negative when they
fell below a certain threshold, say $\tau$. 

Let $p$ represent the estimated probability, and let $\epsilon$
be a small number such that $0<\epsilon<\tau$, where we set $\epsilon=\tau/2$
for implementation. We use the smooth transformation:
\begin{align*}
f(p) & =1(p\geq\tau)\cdot p+1(p<\tau)\cdot\left\{ (\tau-\epsilon)\cdot\tanh\left\{ \dfrac{p-\tau}{\tau-\epsilon}\right\} +\tau\right\} \\
f^{\prime}(p) & =1(p\geq\tau)+1(p<\tau)\cdot\left\{ 1-\tanh^{2}\left\{ \dfrac{p-\tau}{\tau-\epsilon}\right\} \right\} 
\end{align*}
We adopt this transformation because it has a continuous derivative
at $p=\tau$, which is used for gradient calculation. Therefore, the
Step 3 is implemented in practice as:
\begin{align*}
\hat{\rho}_{sy}= & \argmax_{\rho_{sy}}\dfrac{1}{n}\sum_{i=1}^{n}D_{i}\Big[\bar{J}_{i}^{s}\bar{I}_{i}^{y}\log f\left(\Phi_{2}(Z_{i}^{\prime}\hat{\mu}_{s},X_{i}^{\prime}\hat{\nu}_{y};g(Z_{i}^{\prime}\rho_{sy}))\right)\\
 & \;\;\;\;\;\;\;\;\;\;\;\;\;\;\;\;\; +\bar{J}_{i}^{s}I_{i}^{y}\log f\left(\Phi_{2}(Z_{i}^{\prime}\hat{\mu}_{s},-X_{i}^{\prime}\hat{\nu}_{y};-g(Z_{i}^{\prime}\rho_{sy}))\right)\\
 & \;\;\;\;\;\;\;\;\;\;\;\;\;\;\;\;\;+J_{i}^{s}\bar{I}_{i}^{y}\log f\left(\Phi_{2}(Z_{i}^{\prime}\hat{\mu}_{0},X_{i}^{\prime}\hat{\nu}_{y};g(X_{i}^{\prime}\hat{\rho}_{0y}))-\Phi_{2}(Z_{i}^{\prime}\hat{\mu}_{s},X_{i}^{\prime}\hat{\nu}_{y};g(Z_{i}^{\prime}\rho_{sy}))\right)\\
 & \;\;\;\;\;\;\;\;\;\;\;\;\;\;\;\;\;+J_{i}^{s}I_{i}^{y}\log f\left(\Phi_{2}(Z_{i}^{\prime}\hat{\mu}_{0},-X_{i}^{\prime}\hat{\nu}_{y};-g(X_{i}^{\prime}\hat{\rho}_{0y}))-\Phi_{2}(Z_{i}^{\prime}\hat{\mu}_{s},-X_{i}^{\prime}\hat{\nu}_{y};-g(Z_{i}^{\prime}\rho_{sy}))\right)\Big].
\end{align*}

\subsection{Discussion of Assumption \ref{ass:fclt}(4)}\label{app:main-remark-remark2}

The condition on the joint density function imposes smoothness on the coefficients of the BDR because\footnote{For a formal result, see Lemma \ref{lem:conz} in SM \ref{app:notation}.}
$$
f_{S^*,Y^* \mid Z}(s,y \mid z) = \dfrac{\partial \Phi_{2}(-z^{\prime}\mu(s),-z^{\prime}\nu(y);g(z^{\prime}\rho(s,y)))}{\partial s \partial y}.
$$
We focus on the case where both $Y$ and $S$ are continuous because it is the most challenging, but it is easy to accommodate cases where $Y$ and/or $S$ are discrete by suitably modifying Assumption \ref{ass:fclt}(4). For example if $Y$ is continuous and $S$ discrete,  Assumption \ref{ass:fclt}(4) becomes: $\mathcal{Y}$ is a bounded interval,  $\mathcal{S}$
is finite, and the conditional density function $f_{Y^* \mid X}(y \mid x)$ exists, is uniformly bounded above, and is uniformly continuous in $(y,x)$ on $\mathcal{Y}\times \mathcal{X}_1$, where $\mathcal{X}_1$ is the support of $X$ conditional on $S > 0$.  The restrictions that $\mathcal{Y}$ and $\mathcal{S}$ be compact subsets of the supports is imposed to avoid far tail estimation. They can be dispensed by imposing structure on the BDR's parameters at the tails.

\subsection{Monte Carlo Simulation}\label{app:simulation}
We conduct a Monte Carlo simulation calibrated to the empirical application to study the finite-sample performance of our estimator. The data generating process is an HSM with censored selection:
\begin{align*}
    Y^{*} &=X^{\prime}\nu+\sigma_{U}U\\
    S^{*} &= Z^{\prime}\mu+\sigma_{V}V\\
    S &=\max(S^{*},0),\quad Y=Y^{*}\text{ if }S>0,
\end{align*}
where, conditional on $Z$, $(U,V)$ follows a standard bivariate Gaussian distribution with correlation parameter $\tilde{\rho}$. We calibrate this design using data for women from the final ten years of the sample, 2004--2013. The parameters $(\nu,\sigma_{U},\tilde{\rho})$ are estimated using the HSM, and $(\mu,\sigma_{V})$ are estimated using a Tobit model. 

 We generate 500 artificial datasets from the previous design and estimate our distribution regression model with censored selection. We use the same specifications for the selection and outcome equations as in the empirical application and Specification 1 for the selection sorting function, that is, $g(Z^{\prime}\rho(s,y))=g(\rho(s,y))$. We implement Algorithm \ref{alg:est}, where $\overline{\mY}$ is a grid of sample quantiles of log real hourly wage with quantile indices $\lbrace 0.1,0.15,\dots,0.9\rbrace$, calculated from the original subsample used for calibration and $\overline{\mS}=\lbrace 0,34,40\rbrace$. We focus on the estimator of $\rho(s,y)$.\footnote{The simulation analysis in CFL examines the parameters in the outcome and selection equations.} Although $\rho(s,y)$ is constant in the data generating process, we do not impose this restriction in estimation.

\begin{figure} [ht!]
\centering     
\includegraphics[width=\textwidth]{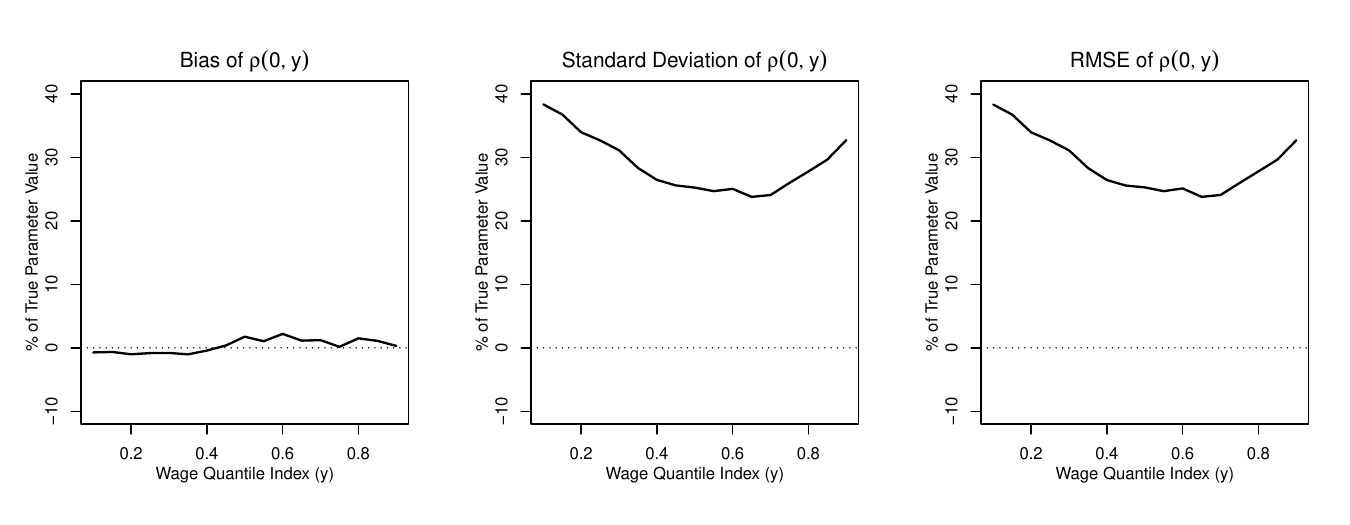}         
\caption{Bias, SD, and RMSE for $\rho(0,y)$} \label{fig_sim1}
\end{figure}

\begin{figure} [ht!]
\centering     
\includegraphics[width=\textwidth]{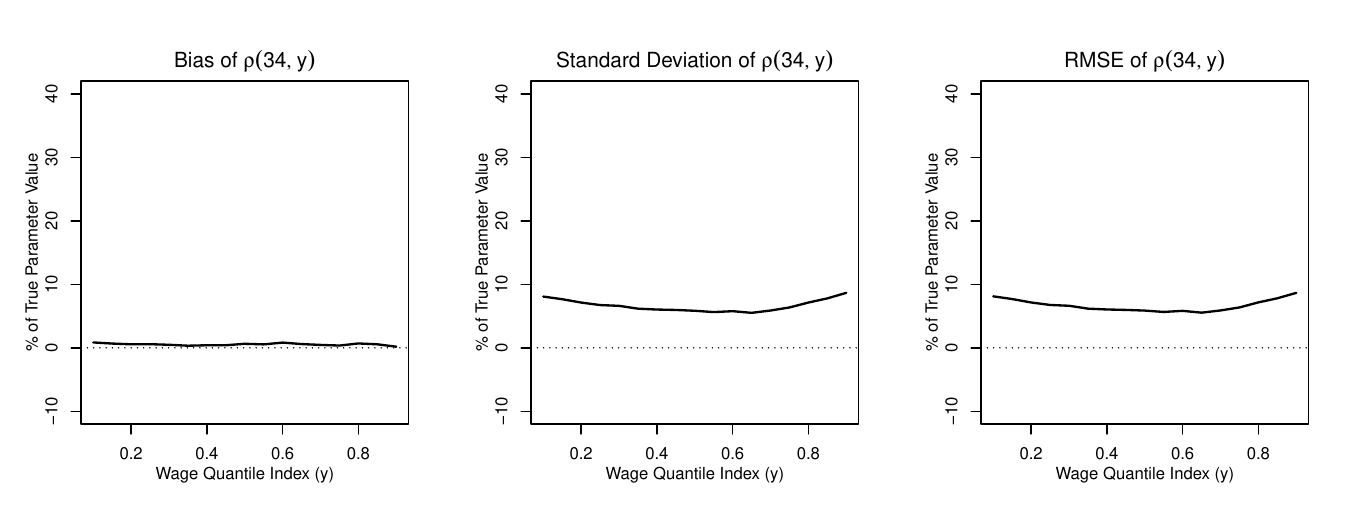}         
\caption{Bias, SD, and RMSE for $\rho(34,y)$} \label{fig_sim2}
\end{figure}

\begin{figure} [ht!]
\centering     
\includegraphics[width=\textwidth]{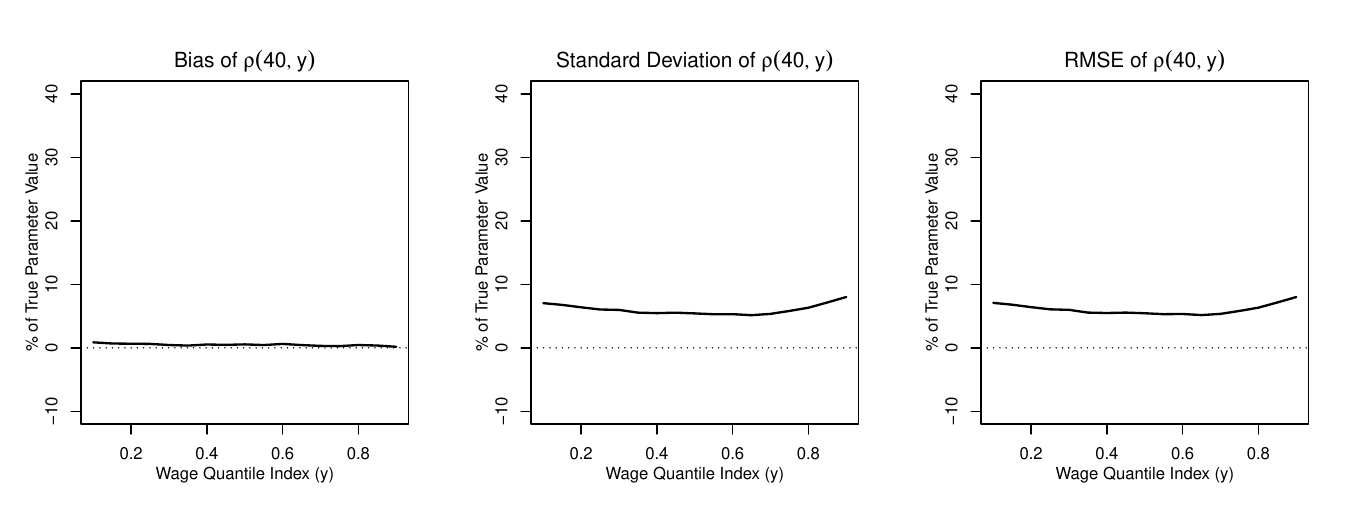}         
\caption{Bias, SD, and RMSE for $\rho(40,y)$} \label{fig_sim3}
\end{figure}

Figures \ref{fig_sim1}, \ref{fig_sim2}, and \ref{fig_sim3} report the bias, standard deviation (SD), and root mean square error (RMSE) of the estimator of $\rho(s,y)$ for $s\in \lbrace 0,34,40\rbrace$. All results are presented as percentages of the true parameter value.  Consistent with the simulation results in CFL, 
Figure \ref{fig_sim1} shows that the biases  for $\rho(0,y)$ are small across quantile indices, while the SDs and RMSEs are larger in the tails than around the median. The results for $\rho(34,y)$ and $\rho(40,y)$ exhibit smaller biases and SDs than those for $\rho(0,y)$, as well as more stable behavior in the tails.

Table \ref{tab:confidence_bands} reports the finite-sample performance of the 95\% confidence bands constructed using Algorithm \ref{alg:mboot}, with $B=200$ bootstrap repetitions and the same grid points $\overline{\mY}$ used for estimation. We report the average length, average estimated critical value, and empirical coverage probability of 95\% confidence bands that are uniform over $y\in \overline{\mY}$, separately for $\rho(0,y)$, $\rho(34,y)$, and $\rho(40,y)$. The ``All'' column reports results for a confidence band that is uniform jointly over $(s,y)\in \overline{\mS}\times \overline{\mY}$. For comparison, we also report the average pointwise coverage probability obtained using the standard normal critical value of 1.96. The final row reports, for each threshold, the average ratio of the estimated standard error to the simulation standard deviation across values of $y\in \overline{\mY}$.

\begin{table}[!htbp]
\centering
\begin{threeparttable}
\caption{Properties of 95\% Confidence Bands}
\label{tab:confidence_bands}
\begin{tabular}{lcccc}
\hline\hline
 & $\rho(0,y)$ & $\rho(34,y)$ & $\rho(40,y)$ & All\\
\hline
Average Band Length              & 0.47 & 0.11 & 0.10 & 0.25\\
Average Critical Value           & 2.76 & 2.76 & 2.76 & 3.03\\
Coverage Uniform Band (\%)       & 93   & 95   & 95   & 93\\
Coverage Pointwise Interval (\%) & 69   & 68   & 71   & 49\\
Average SE/SD                    & 1.00 & 1.02 & 1.03 & \\
\hline\hline
\end{tabular}
\begin{tablenotes}
\footnotesize
\item Notes: Results are based on 500 Monte Carlo replications and 200 bootstrap draws. The first three columns report coverage uniform over $y\in\overline{\mY}$ for each fixed $s\in\overline{\mS}$. The ``All'' column reports joint coverage over $y\in\overline{\mY}$ and all three values of $s$. Pointwise intervals are constructed using the standard normal critical value of 1.96.
\end{tablenotes}

\end{threeparttable}
\end{table}

The uniform confidence bands achieve coverage rates close to the nominal level, whereas the pointwise intervals, as expected, substantially undercover the entire function. The joint confidence band across all three threshold values of $s$ is slightly more conservative, with a larger average critical value of 3.03 than 2.76 for the threshold-specific bands.\footnote{Note that the simulation standard error is $\sqrt{.05*.95/500} \approx 1\%$.} The estimated standard errors based on the asymptotic approximation also provide a reasonably accurate measure of the estimator’s sampling variability, with average SE/SD ratios close to one across all three threshold values.

\section{Supplements to Section \ref{sec:empirics}}
\subsection{Descriptive Statistics}\label{app:B1}

\begin{table}[H]
\centering \caption{Descriptive Statistics of the Sample}
\renewcommand{\arraystretch}{0.9}
\begin{tabular}{l*{6}{c}}
\hline\hline                 &\multicolumn{2}{c}{\textbf{Full}}         &\multicolumn{2}{c}{\textbf{Male}}         &\multicolumn{2}{c}{\textbf{Female}}    \\  
              &     All        &  Employed         &     All        &   Employed         &   All         & Employed \\ 
\hline Log Hourly Wage &                  &    2.38         &                  &    2.54         &                  &    2.21         \\ 
\hline  
Employed        &    0.74         &                  &    0.83         &                  &    0.66        &                  \\ 
Weekly Work Hours  & 26.9  & 36.5 & 35.5 & 42.9 & 19.6 & 29.7 \\ 
Part-time  & 0.19 & 0.26 & 0.04 & 0.05 & 0.33 & 0.50 \\
Full-time  & 0.34 & 0.46 & 0.42 & 0.50 & 0.27 & 0.41 \\
Overtime  & 0.20 & 0.28 & 0.37 & 0.45 & 0.06 & 0.09 \\
\hline  
\multicolumn{7}{l}{Ceased School at} \\ \hspace{3mm} $\leq$ 15       &    0.33         &    0.30         &    0.33         &    0.31         &    0.33         &    0.29         \\ \hspace{3mm}16       &    0.31         &    0.30         &    0.32         &    0.32         &    0.30         &    0.29         \\ \hspace{3mm}17-18     &    0.18         &    0.19         &    0.16         &    0.17         &    0.20         &    0.22         \\ \hspace{3mm}19-20      &   0.04         &   0.05         &   0.04         &   0.04         &   0.04         &   0.05         \\ \hspace{3mm}21-22      &   0.09         &    0.11         &   0.09         &    0.10         &   0.09         &    0.12         \\ \hspace{3mm}$\geq$23     &   0.05         &   0.05         &   0.06         &   0.06         &   0.04         &   0.04         \\ Age            &    40.13         &    39.84         &    40.22         &    39.76         &    40.06         &    39.92         \\ Married        &    0.76         &    0.79         &    0.78         &    0.81         &    0.75         &    0.76         \\ Benefit Income  &    5.44         &    5.50         &    5.25         &    5.29         &    5.60         &    5.73         \\ \hline Observations    &   258,900         &   190,765         &   119,396         &    98,764         &   139,504         &    92,001         \\ \hline\hline 
\end{tabular} 
\label{table1}
\end{table}

\subsection{Selection Sorting Functions for Specifications 3 and 4}\label{app:sorting}
\begin{figure} [H]
\centering     
\includegraphics[width=.85\textwidth, height=9cm]{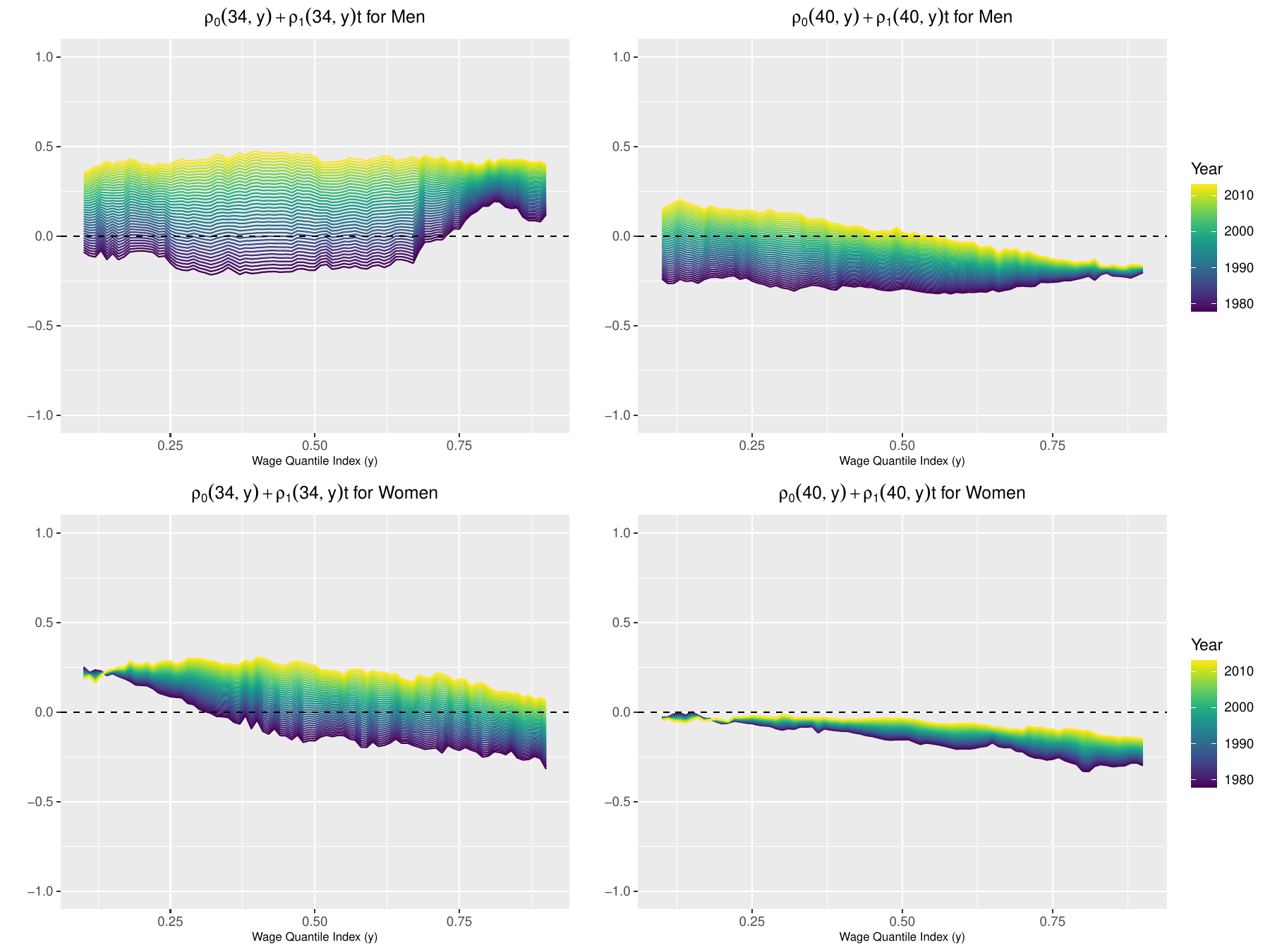}         
\caption{Selection sorting function using Specification 3} \label{fig6}
\end{figure}

\begin{figure} [H]
\centering     
\includegraphics[width=\textwidth, height=9cm]{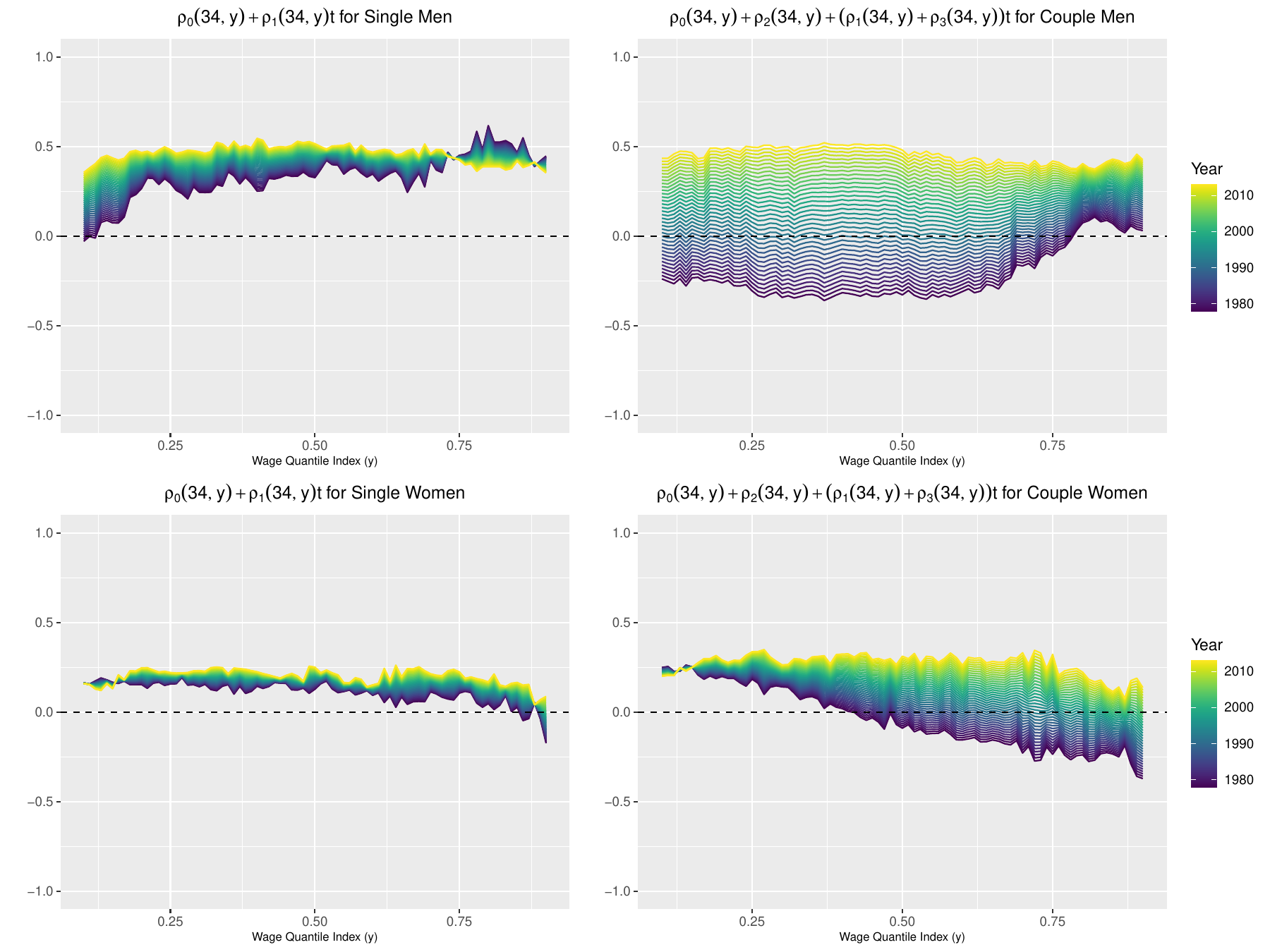}         
\caption{Selection sorting into full-time using Specification 4} \label{fig10}
\end{figure}

\begin{figure} [H]
\centering     
\includegraphics[width=\textwidth, height=9cm]{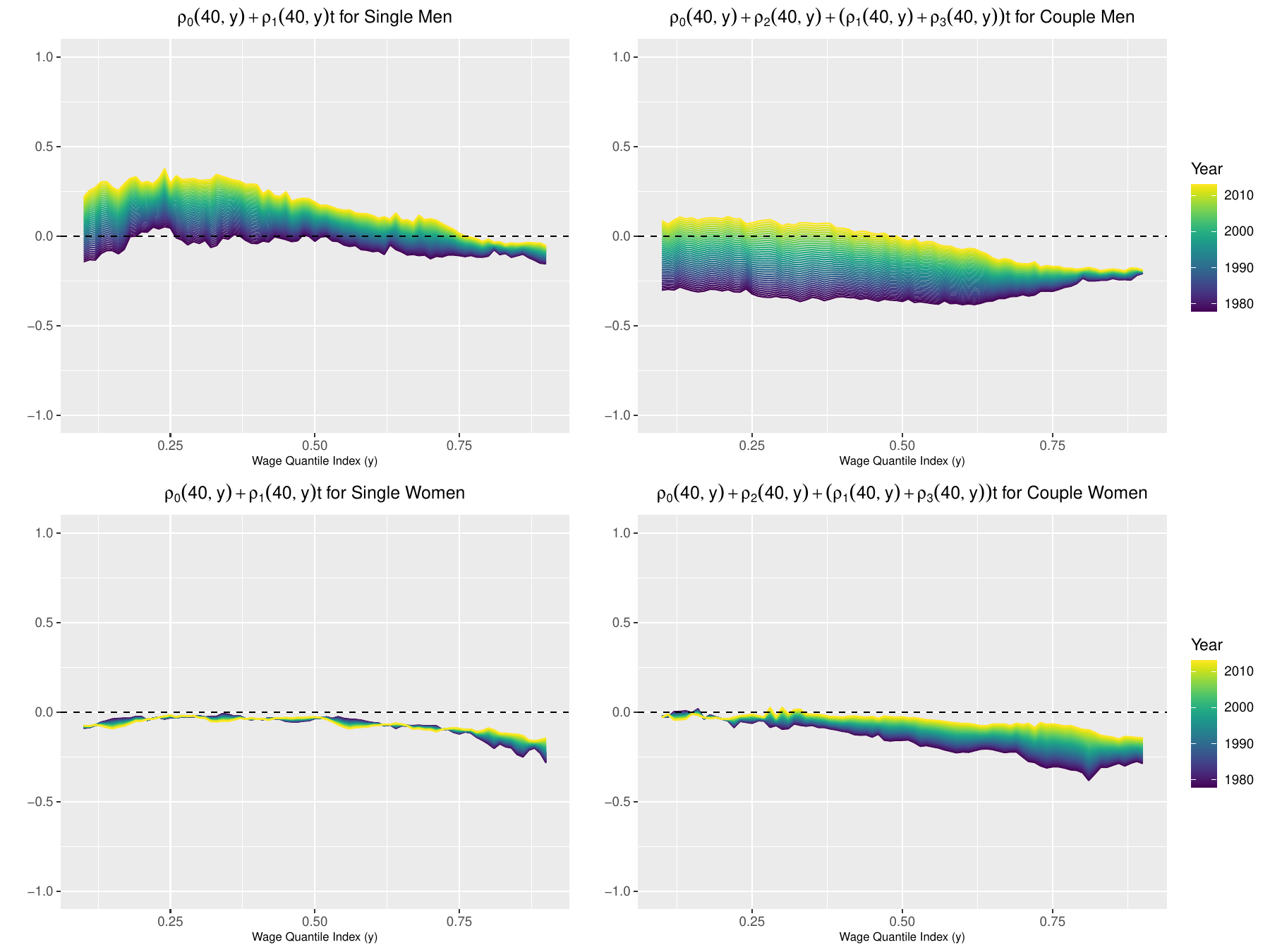}         
\caption{Selection sorting into overtime using Specification 4} \label{fig11}
\end{figure}

\subsection{Work hours decomposition}\label{app:work hours}
Lastly, we decompose the difference in the distribution functions of observed work hours between genders using counterfactual distributions. Let
\begin{align*}
F_{S<r,k>}(s) = \int \Phi(-z^{\prime}\mu_{r}(s)) \mathrm{d} F_{Z_{k}}(z), \quad s \geq 0,
\end{align*}
where $\mu_{r}$ is the coefficient of the selection equation in group
$r$, and $F_{Z_{k}}$ is the distribution of characteristics in group
$k$. We decompose the difference in the distribution functions as
\begin{align*}
    \underset{\text{Total}}{\underbrace{F_{S<0,0>}-F_{S<1,1>}}}=\underset{\text{Structure}}{\underbrace{[F_{S<0,0>}-F_{S<1,0>}]}}+\underset{\text{Composition}}{\underbrace{[F_{S<1,0>}-F_{S<1,1>}]}}.
\end{align*}
The first term is the structure effect, showing differences in how work hours are selected across genders, and the second term is the composition effect.\footnote{Note that we are using distribution functions for work hours decomposition. Therefore, we subtract the women's distribution function from the men's, whereas the order of subtraction is reversed when using quantile functions.} 

The distribution of work hours for men first-order stochastically dominates the distribution for women. The majority of the difference is explained by differences in structure, whereas differences in composition have very little explanatory power. This implies that differences in how men and women choose work hours, such as according to education level, marital status, and number of children, lead to the gender gap in work hours rather than differences in characteristics. This aligns with the wage decomposition result in Section 4.3.2 in that composition effects have little explanatory power in gender wage gap.

\begin{figure}[H]
\centering     
\includegraphics[width=.85\textwidth]{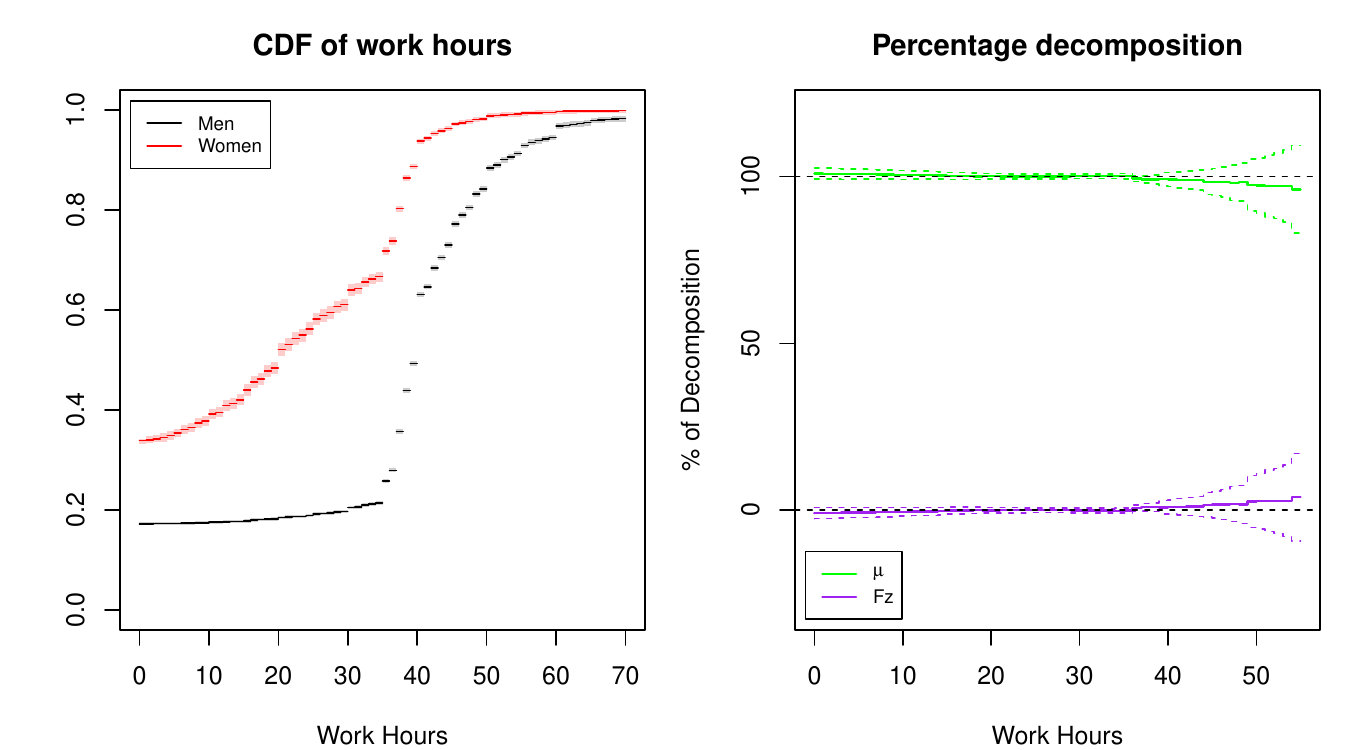}         
\caption{Estimates for the distribution of work hours and decomposition between men and women} \label{fig9}
\end{figure}

\section{Proofs}\label{app:proof}
\subsection{Proof of Theorem \ref{thm:id}} We need to show existence and uniqueness of the solution in $\rho_z(s,y)$ to the nonlinear equation \eqref{eq:rho_id}. Existence of solution follows from Lemma \ref{lemma:lgr} and Assumption \ref{ass:excl}. Uniqueness of solution follows because the left hand side of \eqref{eq:rho_id} does not depend on $\rho_z(s,y)$ and the right hand side of \eqref{eq:rho_id} is strictly increasing in $\rho_z(s,y)$ because
$$
\dfrac{\partial \Phi_{2}(\mu,\nu;\rho)}{\partial \rho} = \phi_{2}(\mu,\nu;\rho) > 0, \quad (\mu,\nu,\rho) \in \mathbb{R}^2 \times [-1,1],
$$
where $\phi_2(\cdot,\cdot;\rho)$ is the joint probability density function of a standard bivariate Gaussian variable with correlation parameter $\rho$. 

\subsection{Notation and Auxiliary Result for Theorems \ref{thm:fclt} and \ref{thm:fclt_boot}}\label{app:notation}
Let 
$$\mathbb{G}_{n}[f]:=\mathbb{G}_{n}[f(W)]:=\dfrac{1}{\sqrt{n}} \sum_{i=1}^{n}(f(W_{i})-\Ep[f(W_{i})]).$$
When the function $\hat{f}$ is estimated, the notation should be interpreted as
$$\mathbb{G}_{n}[\hat{f}]=\mathbb{G}_{n}[f]|_{f=\hat{f}}.$$

We also follow the notation and definitions in \cite{van1996weak}
of bootstrap consistency. Let $D_{n}$ denote the data vector and $E_{n}$ be
the vector of bootstrap weights. Consider the random element $%
Z_{n}^{b}=Z_{n}(D_{n},E_{n})$ in a normed space $\mathbb{Z}$. We say that
the bootstrap law of $Z_{n}^{b}$ consistently estimates the law of some
tight random element $Z$ and write $Z_{n}^{b}\rightsquigarrow_{\Pr}Z$ in $%
\mathbb{Z}$ if 
\begin{equation}
\begin{array}{r}
\sup_{h\in\text{BL}_{1}(\mathbb{Z})}\left|\Ep^{b}h%
\left(Z_{n}^{b}\right)- \Ep h(Z)\right|\rightarrow_{\Pr^{*}}0,%
\end{array}
\label{boot1}
\end{equation}
where $\text{BL}_{1}(\mathbb{Z})$ denotes the space of functions with
Lipschitz norm at most 1, ${\Ep}^{b}$ denotes the conditional
expectation with respect to $E_{n}$ given the data $D_{n}$, and $%
\rightarrow_{\Pr^{*}}$ denotes convergence in (outer) probability.

We use the $Z$-process framework from SM E.1 of \citet{chernozhukov2013inference}. Let $W:=(S,Y,Z)$ denote all the observed variables, $\xi_{sy}=(\mu_{0}^{\prime},\mu_{s}^{\prime},\theta_{y}^{\prime},\rho_{sy}^{\prime})^{\prime}$ be a vector with the model parameters of the first, second, and third steps, and $d_{\xi}:=\text{dim }\xi_{sy}$. Let $\varphi_{sy,\xi}(W)$ be the vector of score functions
\begin{align}\label{eq:scores}
\varphi_{sy,\xi}(W) & :=\begin{bmatrix}S_{10,\xi}(W)\\
S_{1s,\xi}(W)\\
S_{2y,\xi}(W)\\
S_{3sy,\xi}(W)
\end{bmatrix} :=\begin{bmatrix}\frac{\partial\ell_{10,\xi}(W)}{\partial\mu_{0}}\\
\frac{\partial\ell_{1s,\xi}(W)}{\partial\mu_{s}}\\
\frac{\partial\ell_{2y,\xi}(W)}{\partial\theta_{y}}\\
\frac{\partial\ell_{3sy,\xi}(W)}{\partial\rho_{sy}}
\end{bmatrix},
\end{align}
where
\begin{align*}
\ell_{10,\xi}(W) &:=D\log\Phi(Z^{\prime}\mu_{0})+(1-D)\log\Phi(-Z^{\prime}\mu_{0}),\\
\ell_{1s,\xi}(W) &:=\bar{J}^{s}\log\Phi(Z^{\prime}\mu_{s})+J^{s}\log\Phi(-Z^{\prime}\mu_{s}),\\
\ell_{2y,\xi}(W) &:=D\left[\bar{I}^{y}\log\Phi_{2}\left(Z^{\prime}\mu_{0},X^{\prime}\nu_{y};g(X^{\prime}\rho_{0y})\right)+I^{y}\log\Phi_{2}\left(Z^{\prime}\mu_{0},-X^{\prime}\nu_{y};-g(X^{\prime}\rho_{0y})\right)\right],\\
\ell_{3sy,\xi}(W)&=D\Big[\bar{J}^{s}\bar{I}^{y}\log\Phi_{2}\left(Z^{\prime}\mu_{s},X^{\prime}\nu_{y};g(Z^{\prime}\rho_{sy})\right)+\bar{J}^{s}I^{y}\log\Phi_{2}\left(Z^{\prime}\mu_{s},-X^{\prime}\nu_{y};-g(Z^{\prime}\rho_{sy})\right)\\
 & \quad +J^{s}\bar{I}^{y}\log\left\{ \Phi_{2}\left(Z^{\prime}\mu_{0},X^{\prime}\nu_{y};g(X^{\prime}\rho_{0y})\right)-\Phi_{2}\left(Z^{\prime}\mu_{s},X^{\prime}\nu_{y};g(Z^{\prime}\rho_{sy})\right)\right\} \\
 & \quad +J^{s}I^{y}\log\left\{ \Phi_{2}\left(Z^{\prime}\mu_{0},-X^{\prime}\nu_{y};-g(X^{\prime}\rho_{0y})\right)-\Phi_{2}\left(Z^{\prime}\mu_{s},-X^{\prime}\nu_{y};-g(Z^{\prime}\rho_{sy})\right)\right\} \Big],
\end{align*}
are the likelihood functions in each step. Also, let
\[
J(s,y):=\Ep\left[\dfrac{\partial \varphi_{sy,\xi}(W)}{\partial \xi^{\prime}}\right]=\begin{bmatrix}H_{10} & 0 & 0 & 0\\
0 & H_{1s} & 0 & 0\\
J_{2y} & 0 & H_{2y} & 0\\
J_{3sy}^{(1)} & J_{3sy}^{(2)} & J_{3sy}^{(3)} & H_{3sy}
\end{bmatrix},
\]
be the expected Hessian evaluated at the true value of $\xi_{sy}$. Note that
\begin{footnotesize}{\begin{equation}\label{eq:inv-hess}
J^{-1}(s,y)=\begin{bmatrix}H_{10}^{-1} & 0 & 0 & 0\\
0 & H_{1s}^{-1} & 0 & 0\\
-H_{2y}^{-1}J_{2y}H_{10}^{-1} & 0 & H_{2y}^{-1} & 0\\
-H_{3sy}^{-1}\lbrace J_{3sy}^{(1)}-J_{3sy}^{(3)}H_{2y}^{-1}J_{2y} \rbrace H_{10}^{-1} & 
-H_{3sy}^{-1}J_{3sy}^{(2)}H_{1s}^{-1} & 
-H_{3sy}^{-1}J_{3sy}^{(3)}H_{2y}^{-1} & 
H_{3sy}^{-1}
\end{bmatrix},    
\end{equation}}\end{footnotesize}using the inverse of the partitioned matrix formula. Analytical expressions for the components of the score and expected Hessian are given in SM \ref{app:expressions}.

The following lemma from CFL provides sufficient conditions to verify Condition $Z$ in \citet{chernozhukov2013inference}. Let $(\xi,s,y)\mapsto\Psi(\xi,s,y):=\Ep[\varphi_{sy,\xi}(W)].$ 

\begin{lem}[Sufficient Condition for $Z$]\label{lem:conz}
Suppose that $\Xi$ is a compact subset of $\mathbb{R}^{d_{\xi}}$ and $\mathcal{SY}:= \mathcal{S}\times\mathcal{Y}$ is a compact set in $\mathbb{R}^{2}$. Let $\mathcal{I}$ be an open set containing $\mathcal{SY}$. Suppose that (a) $\Psi:\Xi\times\mathcal{I}\mapsto\mathbb{R}^{d_{\xi}}$ is continuous, (b) $\xi\mapsto\Psi(\xi,s,y)$ possesses a unique zero at $\xi_{sy}$
that is in the interior of $\Xi$ for each $(s,y)\in\mathcal{SY}$, (c) $\partial\Psi(\xi,s,y)/\partial(\xi',s,y)$ exists at $(\xi_{sy},s,y)$, and is continuous at $(\xi_{sy},s,y)$ for each
$(s,y)\in\mathcal{SY}$, and $\dot{\Psi}_{\xi_{sy},s,y}=\partial\Psi(\xi,s,y)/\partial\xi^{\prime} |_{\xi_{sy}}$ obeys $\inf_{(s,y)\in\mathcal{SY}}\inf_{||h||=1}\left\Vert \dot{\Psi}_{\xi_{sy},s,y}h\right\Vert >c_{0}>0$. Then Condition Z of \citet{chernozhukov2013inference} holds and $(s,y)\mapsto\xi_{sy}$ is continuously
differentiable.
\end{lem}

\subsection{Proof of Theorem \ref{thm:fclt}} 
The proof follows the same steps as the proof of Theorem C.1 of CFL. Let $\Psi(\xi,s,y):=P\left[\varphi_{sy,\xi}(W)\right]$ and $\hat{\Psi}(\xi,s,y):=P_{n}\left[\varphi_{sy,\xi}(W)\right]$, where $P_{n}$ represents the empirical measure and $P$ represents the corresponding probability measure. From the first-order conditions, the three-step
estimator satisfies $\hat{\xi}_{sy}=\phi\left(\hat{\Psi}(\cdot,s,y),0\right)$
for each $(s,y)\in\mathcal{SY}$, where $\phi$ is the $Z$-map defined
in SM E.1 of \citet{chernozhukov2013inference}. The random vector $\hat{\xi}_{sy}$ is the estimator of $\xi_{sy}=\phi\left(\Psi(\cdot,s,y),0\right)$ in the notation of this framework. Applying Step 1 below, we obtain
\[
\sqrt{n}\left(\hat{\Psi}-\Psi\right)\rightsquigarrow Z_{\Psi}\text{ in }\ell^{\infty}\left(\mathcal{SY}\times\mathbb{R}^{d_{\xi}}\right)^{d_{\xi}},
\]
where $Z_{\Psi}(s,y,\xi)=\mathbb{G}\varphi_{sy,\xi}$, $\mathbb{G}$ is a $P$-Brownian bridge, and $Z_{\Psi}$ has continuous paths a.s.

Step 2 verifies the conditions of Lemma \ref{lem:conz} for $\dot{\Psi}(\xi_{sy},s,y)=J(s,y)$,
which also implies that $(s,y)\mapsto\xi_{sy}$ is continuously differentiable
in the set $\mathcal{SY}$. Then, by Lemma E.2 of \citet{chernozhukov2013inference}, the
map $\phi$ is Hadamard differentiable with derivative map $(\psi,0)\mapsto-J^{-1}\psi$
at $(\Psi,0)$. Therefore, we can conclude by the functional delta
method that
\begin{align}\label{eq:fclt}
\sqrt{n}\left(\hat{\xi}_{sy}-\xi_{sy}\right)\rightsquigarrow Z_{\xi_{sy}}:=-J^{-1}(s,y)Z_{\Psi}(s,y,\xi_{sy})\text{ in }\ell^{\infty}\left(\mathcal{SY}\right)^{d_{\xi}},
\end{align}
where $(s,y)\mapsto Z_{\xi_{sy}}$ has continuous paths a.s. 

\begin{step}[Donskerness] We verify that $\mathcal{G}=\left\{ \varphi_{sy,\xi}(W):(s,y,\xi)\in\mathcal{SY}\times\mathbb{R}^{d_{\xi}}\right\} $
is a $P$-Donsker with a square-integrable envelope. By examining the expression of 
\[
\varphi_{sy,\xi}(W)=\left[S_{10,\xi}(W)^{\prime},S_{1s,\xi}(W)^{\prime},S_{2y,\xi}(W)^{\prime},S_{3sy,\xi}(W)^{\prime}\right]^{\prime}
\]
in SM \ref{app:expressions}, we see that $\varphi_{sy,\xi}(W)$ is a Lipschitz transformation of VC functions with Lipschitz coefficient bounded by $c\left\Vert Z\right\Vert $ for some constant $c$ and envelope function $c\left\Vert Z\right\Vert $,
which is square-integrable. Hence $\mathcal{G}$ is $P$-Donsker by Example 19.9 in \citet{van1998asymptotic}.
\end{step}

\begin{step}[Verification of the Conditions of Lemma \ref{lem:conz}]
Conditions (a) and (b) follow directly from Assumption \ref{ass:fclt}. To verify (c),
note that for $(\tilde{\xi}_{sy},\tilde{s},\tilde{y})$ in the neighborhood
of $(\xi_{sy},s,y)$,
\[
\dfrac{\partial\Psi(\tilde{\xi}_{sy},\tilde{s},\tilde{y})}{\partial(\tilde{\xi}_{sy}',\tilde{s},\tilde{y})}=\left[J(\tilde{\xi}_{sy},\tilde{s},\tilde{y}),R_{s}(\tilde{\xi}_{sy},\tilde{s},\tilde{y}),R_{y}(\tilde{\xi}_{sy},\tilde{s},\tilde{y})\right].
\]
To characterize the analytical expressions of  $J(\tilde{\xi}_{sy},\tilde{s},\tilde{y})$, $R_{s}(\tilde{\xi}_{sy},\tilde{s},\tilde{y})$, $R_{y}(\tilde{\xi}_{sy},\tilde{s},\tilde{y})$, we use the notation introduced in SM \ref{app:expressions}. Let $(\tilde{A}_1,\tilde{A}_2,\tilde{A}_3,\tilde{A}_4)$ and $(\hat{A}_1,\hat{A}_2,\hat{A}_3,\hat{A}_4)$ be the same expressions as $(A_1,A_2,A_3,A_4)$ in \eqref{eq:As} but evaluated at $\tilde{\xi}_{sy}=(\tilde{\mu}_{0},\tilde{\mu}_{s},\tilde{\theta}_{y},\tilde{\rho}_{sy})$ and $\xi_{\tilde s \tilde y}$, respectively.\footnote{Note that $\xi_{\tilde s \tilde y}$ is the true value of $\xi_{sy}$ at $(s,y) = (\tilde s, \tilde y)$, whereas $\tilde{\xi}_{sy}$ is a value near $\xi_{sy}$.} 
Then,
\[
R_{s}(\tilde{\xi}_{sy},\tilde{s},\tilde{y}):=\dfrac{\partial\Psi(\tilde{\xi}_{sy},\tilde{s},\tilde{y})}{\partial\tilde{s}}=\Ep\begin{bmatrix}0\\
-f_{S^*\mid Z}(\tilde{s}\mid Z)G_{1}(Z^{\prime}\tilde{\mu}_{s})Z\\
0\\
R_{s}^{(4)}(\tilde{\xi}_{sy},\tilde{s},\tilde{y})
\end{bmatrix},
\]
where 
\begin{align*}
R_{s}^{(4)}(\tilde{\xi}_{sy},\tilde{s},\tilde{y})  := &-\left\{ \dfrac{1}{\tilde{A}_1}+\dfrac{1}{\tilde{A}_3}\right\}  f_{S^* \mid Z}(\tilde{s}\mid  Z)\Phi_{2,\tilde{\xi}_{sy}}^{\rho}(Z)\dot g(Z^{\prime}\tilde{\rho}_{sy}) Z\\
 & +\left\{ \dfrac{1}{\tilde{A}_1} + \dfrac{1}{\tilde{A}_2} +\dfrac{1}{\tilde{A}_3} +\dfrac{1}{\tilde{A}_4}\right\} \Phi_{2,\xi_{\tilde s \tilde y}}^{\mu}(Z) \Phi_{2,\tilde{\xi}_{sy}}^{\rho}(Z)\dot g(Z^{\prime}\tilde{\rho}_{sy}) Z;
\end{align*}
and
\[
R_{y}(\tilde{\xi}_{sy},\tilde{s},\tilde{y}) :=\dfrac{\partial\Psi(\tilde{\xi}_{sy},\tilde{s},\tilde{y})}{\partial\tilde{y}}=\Ep\begin{bmatrix}0\\
0\\
R_{y}^{(3)}(\tilde{\xi}_{sy},\tilde{s},\tilde{y}) \\
R_{y}^{(4)}(\tilde{\xi}_{sy},\tilde{s},\tilde{y})
\end{bmatrix},
\]
where
$$
R_{y}^{(3)}(\tilde{\xi}_{sy},\tilde{s},\tilde{y}) := [\Phi_{2,\xi_{\tilde s \tilde y}}^{\nu}(Z)-f_{Y^* \mid X}\left(\tilde{y} \mid x\right) ]\Phi(Z^{\prime}\tilde{\mu}_{0})\begin{bmatrix}G_{2,\tilde{\xi}_{sy}}(Z)\\
\dot g(X^{\prime}\tilde{\rho}_{0y})G_{3,\tilde{\xi}_{sy}}(Z)
\end{bmatrix}\otimes X,
$$
and
\begin{footnotesize}{\begin{align*}
R_{y}^{(4)}(\tilde{\xi}_{sy},\tilde{s},\tilde{y})  := &\left\{ \dfrac{1}{\tilde{A}_1}+\dfrac{1}{\tilde{A}_2}\right\}  [\Phi_{2,\xi_{\tilde s \tilde y}}^{\nu}(Z)] -f_{Y^* \mid X}(\tilde y \mid x)  \Phi_{2,\tilde{\xi}_{sy}}^{\rho}(Z)\dot g(Z^{\prime}\tilde{\rho}_{sy}) Z\\
 & +\left\{ \dfrac{1}{\tilde{A}_3}+\dfrac{1}{\tilde{A}_4}\right\} [ \Phi_{2,\xi_{\tilde s \tilde y}}^{\nu}(Z) -  \Phi_{2,\xi_{0 \tilde y}}^{\nu}(Z)] \Phi_{2,\tilde{\xi}_{sy}}^{\rho}(Z)\dot g(Z^{\prime}\tilde{\rho}_{sy}) Z.
\end{align*}}\end{footnotesize}

Also, 
\[
J(\tilde{\xi}_{sy},\tilde{s},\tilde{y}):=\dfrac{\partial\Psi(\tilde y,\tilde{s},\tilde{\xi}_{sy})}{\partial \tilde{\xi}_{sy}'}=\begin{bmatrix}J_{11}(\tilde{\xi}_{sy},\tilde{s},\tilde{y}) & 0 & 0 & 0\\
0 & J_{22}(\tilde{\xi}_{sy},\tilde{s},\tilde{y}) & 0 & 0\\
J_{31}(\tilde{\xi}_{sy},\tilde{s},\tilde{y}) & 0 & J_{33}(\tilde{\xi}_{sy},\tilde{s},\tilde{y}) & 0\\
J_{41}(\tilde{\xi}_{sy},\tilde{s},\tilde{y}) & J_{42}(\tilde{\xi}_{sy},\tilde{s},\tilde{y}) & J_{43}(\tilde{\xi}_{sy},\tilde{s},\tilde{y}) & J_{44}(\tilde{\xi}_{sy},\tilde{s},\tilde{y})
\end{bmatrix},
\]
where
\begin{align*}
J_{11}(\tilde{\xi}_{sy},\tilde{s},\tilde{y})&:=\Ep\left[\left\{ \dot G_{1}(Z^{\prime}\tilde{\mu}_{0})\left[D-\Phi(Z^{\prime}\tilde{\mu}_{0})\right]-G_{1}(Z^{\prime}\tilde{\mu}_{0})\dot \Phi(Z^{\prime}\tilde{\mu}_{0})\right\} ZZ^{\prime}\right],\\
J_{22}(\tilde{\xi}_{sy},\tilde{s},\tilde{y})&:=\Ep\left[\left\{ \dot G_{1}(Z^{\prime}\tilde{\mu}_{s})\left[\Phi(Z^{\prime}\mu_{\tilde s})-\Phi(Z^{\prime}\tilde{\mu}_{s})\right]-G_{1}(Z^{\prime}\tilde{\mu}_{s})\dot \Phi(Z^{\prime}\tilde{\mu}_{s})\right\} ZZ^{\prime}\right],
\end{align*}
\begin{align*}
& J_{31}(\tilde{\xi}_{sy},\tilde{s},\tilde{y})  :=\Ep\begin{bmatrix}\left\{ \Phi(Z^{\prime}\mu_{0})\Phi_{2,\tilde{\xi}_{0y}}^{\mu}(Z)-\Phi_{2,\xi_{0\tilde{y}}}(Z)\dot \Phi(Z^{\prime}\tilde{\mu}_{0})\right\} \begin{bmatrix}G_{2,\tilde{\xi}_{0y}}(Z)\\
\dot g(X^{\prime}\tilde{\rho}_{0y})G_{3,\tilde{\xi}_{0y}}(Z)
\end{bmatrix}\otimes XZ^{\prime}\end{bmatrix}\\
 &\quad  +\Ep\begin{bmatrix}\left\{ \Phi(Z^{\prime}\mu_{0})\Phi_{2,\tilde{\xi}_{0y}}(Z)-\Phi_{2,\xi_{0\tilde{y}}}(Z)\Phi(Z^{\prime}\tilde{\mu}_{0})\right\} \begin{bmatrix}G_{2,\tilde{\xi}_{0y}}^{\mu}(Z)\\
\dot g(X^{\prime}\tilde{\rho}_{0y})G_{3,\tilde{\xi}_{0y}}^{\mu}(Z)
\end{bmatrix}\otimes XZ^{\prime}\end{bmatrix},\\
\end{align*}
\begin{footnotesize}{\begin{align*}
&J_{33}  :=-\Ep\left[\Phi(Z^{\prime}\mu_{0})\begin{bmatrix}\Phi_{2,\tilde{\xi}_{0y}}^{\nu}(Z)G_{2,\tilde{\xi}_{0y}}(Z) & \Phi_{2,\tilde{\xi}_{0y}}^{\rho}(Z)G_{2,\tilde{\xi}_{0y}}(Z)\\
\Phi_{2,\tilde{\xi}_{0y}}^{\nu}(Z)\dot g(X^{\prime}\tilde{\rho}_{0y})G_{3,\tilde{\xi}_{0y}}(Z) & \Phi_{2,\tilde{\xi}_{0y}}^{\rho}(Z)\dot g(X^{\prime}\tilde{\rho}_{0y})G_{3,\tilde{\xi}_{0y}}(Z)
\end{bmatrix}\otimes XX^{\prime}\right]\\
 & -\Ep\left[\left\{ \Phi(Z^{\prime}\mu_{0})\Phi_{2,\tilde{\xi}_{0y}}(Z)-\Phi_{2,\xi_{0\tilde{y}}}(Z)\Phi(Z^{\prime}\tilde{\mu}_{0})\right\} \begin{bmatrix}G_{2,\tilde{\xi}_{0y}}^{\nu}(Z) & G_{2,\tilde{\xi}_{0y}}^{\rho}(Z)\\
\dot g(X^{\prime}\tilde{\rho}_{0y})G_{3,\tilde{\xi}_{0y}}^{\nu}(Z) & \dot g(X^{\prime}\tilde{\rho}_{0y})G_{3,\tilde{\xi}_{0y}}^{\rho}(Z)+\ddot g(X^{\prime}\tilde{\rho}_{0y})G_{3,\tilde{\xi}_{0y}}(Z)
\end{bmatrix}\otimes XX^{\prime}\right].
\end{align*}}\end{footnotesize}
Let $G_{4,\xi_{sy}}^{a b}=\Phi_{2}^{ab}(Z^{\prime}\mu_s,-X^{\prime}\nu_y;-g(Z^{\prime}\rho_{sy}))$, where $\Phi_{2}^{ab}(\mu,\nu;\rho):=\partial^2 \Phi_2(\mu,\nu;\rho)/\partial a\partial b$. Then,
\begin{align*}
J_{41}&:=\Ep\left[\left\{ \dfrac{\hat A_3}{\tilde{A}_3^{2}}\left(\dot \Phi(Z^{\prime}\tilde{\mu}_{0})-\Phi_{2,\tilde{\xi}_{0y}}^{\mu}(Z)\right)-\dfrac{\hat  A_4}{\tilde{A}_4^{2}}\Phi_{2,\tilde{\xi}_{0y}}^{\mu}(Z)\right\} \Phi_{2,\tilde{\xi}_{sy}}^{\rho}(Z)\dot g(Z^{\prime}\tilde{\rho}_{sy})  ZZ^{\prime}\right],\\
\end{align*}
\begin{align*}
J_{42}  = &\Ep\Bigg[\left\{ \left(\dfrac{\hat A_1}{\tilde{A}_1^{2}}+\dfrac{\hat A_2}{\tilde{A}_2^{2}}+\dfrac{\hat A_3}{\tilde{A}_3^{2}}+\dfrac{\hat A_4}{\tilde{A}_4^{2}}\right)\Phi_{2,\tilde{\xi}_{sy}}^{\mu}(Z)-\left(\dfrac{\hat A_1}{\tilde{A}_1^{2}}+\dfrac{\hat A_3}{\tilde{A}_3^{2}}\right)\dot \Phi(Z^{\prime}\tilde{\mu}_{s})\right\} \Phi_{2,\tilde{\xi}_{sy}}^{\rho}(Z)\dot g(Z^{\prime}\tilde{\rho}_{sy})  ZZ^{\prime}\\
 & +\left( \dfrac{\hat A_1}{\tilde{A}_1}-\dfrac{\hat A_2}{\tilde{A}_2}-\dfrac{\hat A_3}{\tilde{A}_3}+\dfrac{\hat A_4}{\tilde{A}_4}\right) G_{4,\tilde{\xi}_{sy}}^{\rho\mu}(Z)\dot g(Z^{\prime}\tilde{\rho}_{sy})  ZZ^{\prime}\Bigg],\\
\end{align*}
\begin{footnotesize}{\begin{align*}
J_{43}  = &\Ep\Bigg[\begin{bmatrix}\left( -\dfrac{\hat A_1}{\tilde{A}_1^{2}}-\dfrac{\hat A_2}{\tilde{A}_2^{2}}-\dfrac{\hat A_3}{\tilde{A}_3^{2}}-\dfrac{\hat A_4}{\tilde{A}_4^{2}}\right) \Phi_{2,\tilde{\xi}_{sy}}^{\nu}(Z)+\left( \dfrac{\hat A_3}{\tilde{A}_3^{2}}+\dfrac{\hat A_4}{\tilde{A}_4^{2}}\right) \Phi_{2,\tilde{\xi}_{0y}}^{\nu}(Z)\\
\left( \dfrac{\hat A_3}{\tilde{A}_3^{2}}+\dfrac{\hat A_4}{\tilde{A}_4^{2}}\right) \Phi_{2,\tilde{\xi}_{0y}}^{\rho}(Z)
\end{bmatrix}\Phi_{2,\tilde{\xi}_{sy}}^{\rho}(Z)\dot g(Z^{\prime}\tilde{\rho}_{sy})\otimes ZX^{\prime}\\
 & +\left( \dfrac{\hat A_1}{\tilde{A}_1}-\dfrac{\hat A_2}{\tilde{A}_2}-\dfrac{\hat A_3}{\tilde{A}_3}+\dfrac{\hat A_4}{\tilde{A}_4}\right) \begin{bmatrix}G_{4,\tilde{\xi}_{sy}}^{\rho\nu}(Z)\\
0
\end{bmatrix}\dot g(Z^{\prime}\tilde{\rho}_{sy})\otimes ZX^{\prime}\Bigg],\\
\end{align*}}\end{footnotesize}
\begin{align*}
J_{44} =& \Ep \Bigg[\left\{ \left(-\dfrac{\hat A_1}{\tilde{A}_1^{2}}-\dfrac{\hat A_2}{\tilde{A}_2^{2}}-\dfrac{\hat A_3}{\tilde{A}_3^{2}}-\dfrac{\hat A_4}{\tilde{A}_4^{2}}\right)\Phi_{2,\tilde{\xi}_{sy}}^{\rho}(Z)^{2}\dot g(Z^{\prime}\tilde{\rho}_{sy})  ZZ^{\prime}\right\} \\
 & +\left(\dfrac{\hat A_1}{\tilde{A}_1}-\dfrac{\hat A_2}{\tilde{A}_2}-\dfrac{\hat A_3}{\tilde{A}_3}+\dfrac{\hat A_4}{\tilde{A}_4}\right)\left\{ G_{4,\tilde{\xi}_{sy}}^{\rho\rho}(Z)\dot g(Z^{\prime}\tilde{\rho}_{sy})+\Phi_{2,\tilde{\xi}_{sy}}^{\rho}(Z)\ddot g(Z^{\prime}\tilde{\rho}_{sy})\right\}   ZZ^{\prime}\Bigg].
\end{align*}

We want to show that $(\tilde{\xi}_{sy},\tilde{s},\tilde{y})\mapsto R_{s}(\tilde{\xi}_{sy},\tilde{s},\tilde{y})$, $(\tilde{\xi}_{sy},\tilde{s},\tilde{y})\mapsto R_{y}(\tilde{\xi}_{sy},\tilde{s},\tilde{y})$, and $(\tilde{\xi}_{sy},\tilde{s},\tilde{y})\mapsto J(\tilde{\xi}_{sy},\tilde{s},\tilde{y})$ are continuous at $(\xi_{sy},s,y)$ for each $(s,y)\in\mathcal{SY}$. The verification of the continuity follows from using the expressions, the dominated convergence theorem, and the following ingredients: (1) a.s. continuity of the map $(\tilde{\xi}_{sy},\tilde{s},\tilde{y})\mapsto\partial\varphi_{\tilde{s}\tilde{y},\tilde{\xi}_{sy}}(W)/\partial\tilde{\xi}_{sy}^{\prime}$, (2) domination of $\left\Vert \partial\varphi_{sy,\xi}(W)/\partial\xi^{\prime}\right\Vert$  by a square-integrable function $c\left\Vert Z\right\Vert$  for some constant $c>0$, (3) a.s. continuity and uniform boundedness of the conditional density functions $s\mapsto f_{S^* \mid Z}(s \mid z)$ and $y \mapsto f_{Y^* \mid X}(y \mid x)$ by Assumption \ref{ass:fclt}, and (4) $(\tilde{A}_1,\tilde{A}_2,\tilde{A}_3,\tilde{A}_4)$ being bounded uniformly on $\tilde{\xi}_{sy}\in\mathbb{R}^{d_{\xi}}$ a.s. Finally, the last part of (c) follows because the minimum eigenvalue of $J(s,y)=J(\xi_{sy},s,y)$ is bounded away from zero uniformly on $(s,y)\in\mathcal{SY}$ by Assumption \ref{ass:fclt}.
\end{step}

Finally,  $\hat{\Sigma}_{\rho}(s,y)$ is consistent for $\Sigma_{\rho}(s,y)$ uniformly on $(s,y) \in \mathcal{SY}$ by \eqref{eq:fclt} and the continuous mapping theorem.

\subsection{Proof of Theorem \ref{thm:fclt_boot}}

Let $\hat \psi_{sy,\hat \xi}(W_i) := (\hat \psi_{12sy,\hat \xi}(W_i)', \hat \psi_{3sy,\hat \xi}(W_i)')'$, where $\hat \psi_{12sy,\hat \xi}(W_i)$ and $\hat \psi_{3sy,\hat \xi}(W_i)$ are defined in \eqref{eq:influence12} and \eqref{eq:influence}, be the estimator of $\psi_{sy,\xi}(W_i)$, the influence function of $\hat \xi_{sy}$ given in \eqref{eq:psi}; and $\hat{\xi}_{sy}^{b}:=(\hat{\mu}_{0}^{b\prime},\hat{\mu}_{s}^{b\prime},\hat{\theta}_{y}^{b\prime},\hat{\rho}_{sy}^{b\prime})^{\prime}$ . By the
definition of the multiplier bootstrap draw for the estimator, 
\[
\sqrt{n}(\hat{\xi}_{sy}^{b}-\hat{\xi}_{sy})=\mathbb{G}_{n}\omega^{b}\hat \psi_{sy,\hat \xi} =\mathbb{G}_{n}\omega^{b} \psi_{sy,\xi} +r_{sy},
\]
where $\omega^{b}\sim N(0,1)$ independently of the data by Assumption \ref{ass:mb} and $r_{sy}=\mathbb{G}_{n}\omega^{b}(\hat \psi_{sy,\hat \xi}-\psi_{sy,\xi})$.
Then the result follows from $\mathbb{G}_{n}\omega^{b} \psi_{sy, \xi}\rightsquigarrow_{\Pr}Z_{\xi_{sy}}$
in Step 3 and $r_{sy}\rightsquigarrow_{\Pr}0$ in Step 4.

\begin{step} Recall that $ \psi_{sy, \xi} = -J^{-1} \varphi_{sy,\xi}$ and $\varphi_{sy,\xi}$ is $P$-Donsker as established in Step 1
of the proof of Theorem \ref{thm:fclt}. Then, by $\Ep[\omega^{b}]=0$, $\Ep[(\omega^{b})^{2}]=1$
and the Conditional Multiplier Functional Central Limit Theorem, we obtain
\[
\mathbb{G}_{n}\omega^{b}\psi_{sy, \xi} \rightsquigarrow_{P}Z_{\xi_{sy}} \text{ in }\ell^{\infty}\left(\mathcal{SY}\right)^{d_{\xi}},
\]
where $Z_{\xi_{sy}}$ is the same limit process as in (\ref{eq:fclt}).
\end{step}

\begin{step}
We can express $\hat \psi_{sy,\hat \xi} = \hat J^{-1} \varphi_{sy,\hat \xi}$, with $\hat J(s,y) := \sum_{i=1}^n \partial_{\xi'} \varphi_{sy,\hat \xi}(W_i)$.  Note that $r_{sy}\rightsquigarrow0$ follows from $\varphi_{sy,\hat \xi} - \varphi_{sy,\xi} \rightsquigarrow 0$ and $\hat J - J \rightsquigarrow0$, because $ \varphi_{sy,\xi}$ is $P$-Donsker,  $\sqrt{n}(\hat{\xi}_{sy}-\xi_{sy})=O_{p}(1)$
uniformly in $(s,y)\in\mathcal{SY}$ by Theorem \ref{thm:fclt}, and 
the continuous mapping theorem. To show that $r_{sy}\rightsquigarrow_{\Pr}0$, note that 
this statement means that for any $\epsilon>0$, $\Ep^{b}1\left(\left\Vert r_{sy}\right\Vert _{2}>\epsilon\right)=o_{\Pr}(1)$
uniformly in $(s,y)\in\mathcal{SY}$. Then, the result follows from
the Markov inequality and 
\[
\Ep\left[\Ep^{b}1\left(\left\Vert r_{sy}\right\Vert _{2}>\epsilon\right)\right]=P\left(\left\Vert r_{sy}\right\Vert _{2}>\epsilon\right)=o(1)
\]
uniformly in $(s,y)\in\mathcal{SY}$, where the latter holds by the
Law of Iterated Expectations and $r_{sy}\rightsquigarrow0$.
\end{step}

\subsection{Asymptotic Variance}\label{app:expressions}
We provide analytical expressions for the score, expected Hessian and influence function of the estimator. We use these expressions to construct an estimator of the asymptotic variance
\subsubsection{Analytical Expressions} Let $\theta_{y}=(\nu_{y},\rho_{0y})$, $\eta_{sy} = (\mu_0,\mu_s,\theta_y)$ and $\xi_{sy}=(\eta_{sy},\rho_{sy})$. In Algorithm \ref{alg:est}, the likelihood functions in each step evaluated at the true parameter values are
\begin{footnotesize}{\begin{eqnarray*}\label{eq:obj}
L_{1}(\mu_{s}) &:=& \dfrac{1}{n}\sum_{i=1}^{n} \ell_{1s,\xi}(W_i) = \dfrac{1}{n}\sum_{i=1}^{n}\left[\bar{J}_{i}^{s}\log\Phi(Z_{i}^{\prime}\mu_{s})+J_{i}^{s}\log\Phi(-Z_{i}^{\prime}\mu_{s})\right], \\
L_{2}(\theta_{y},\mu_{0})&:=& \dfrac{1}{n}\sum_{i=1}^{n} \ell_{2y,\xi}(W_i) 
\\ &=&  \dfrac{1}{n}\sum_{i=1}^{n}D_{i}\left[\bar{I}_{i}^{y}\log\Phi_{2}(Z_{i}^{\prime}\mu_{0},X_{i}^{\prime}\nu_{y};g(X_{i}^{\prime}\rho_{0y}))+I_{i}^{y}\log\Phi_{2}(Z_{i}^{\prime}\mu_{0},-X_{i}^{\prime}\nu_{y};-g(X_{i}^{\prime}\rho_{0y}))\right], \notag \\
L_{3}(\rho_{sy},\eta_{sy}) &:=& \dfrac{1}{n}\sum_{i=1}^{n} \ell_{3sy,\xi}(W_i)
\\ &=& \dfrac{1}{n}\sum_{i=1}^{n} D_{i}\Big[\bar{J}_{i}^{s}\bar{I}_{i}^{y}\log\Phi_{2}(Z_{i}^{\prime}\mu_{s},X_{i}^{\prime}\nu_{y};g(Z_{i}^{\prime}\rho_{sy}))+
\bar{J}_{i}^{s}I_{i}^{y}\log\Phi_{2}(Z_{i}^{\prime}\mu_{s},-X_{i}^{\prime}\nu_{y};-g(Z_{i}^{\prime}\rho_{sy})) \notag \\
 & &\;\;\;\;\;\;\;\;\;\;\;\;+J_{i}^{s}\bar{I}_{i}^{y}\log\left\{ \Phi_{2}(Z_{i}^{\prime}\mu_{0},X_{i}^{\prime}\nu_{y};g(X_{i}^{\prime}\rho_{0y}))-\Phi_{2}(Z_{i}^{\prime}\mu_{s},X_{i}^{\prime}\nu_{y};g(Z_{i}^{\prime}\rho_{sy}))\right\} \notag \\
 & &\;\;\;\;\;\;\;\;\;\;\;\;+J_{i}^{s}I_{i}^{y}\log\left\{ \Phi_{2}(Z_{i}^{\prime}\mu_{0},-X_{i}^{\prime}\nu_{y};-g(X_{i}^{\prime}\rho_{0y}))-\Phi_{2}(Z_{i}^{\prime}\mu_{s},-X_{i}^{\prime}\nu_{y};-g(Z_{i}^{\prime}\rho_{sy}))\right\} \Big].\notag
\end{eqnarray*}}\end{footnotesize}

It is convenient to introduce some notation to simplify the analytical expressions. Let
$\Phi_{2,\xi_{sy}}(Z):=\Phi_{2}\left(Z^{\prime}\mu_{s},X^{\prime}\nu_{y};g(Z^{\prime}\rho_{sy})\right)$, $\Phi_{2,\xi_{sy}}^{a}(Z):=\Phi_{2}^{a}\left(Z^{\prime}\mu_{s},X^{\prime}\nu_{y};g(Z^{\prime}\rho_{sy})\right)$, 
where $\Phi_{2}^{a}(\mu,\nu;\rho):= \partial \Phi_{2}(\mu,\nu;\rho)/\partial a$. Also, let $G_{1}(u) := \dot \Phi(u)/[\Phi(u)\Phi(-u)]$,
\begin{align*}
    &
    G_{2}(\mu,\nu;\rho):=\dfrac{\Phi_{2}^{\nu}(\mu,\nu;\rho)}{\Phi_{2}(\mu,\nu;\rho)[\Phi(\mu)-\Phi_{2}(\mu,\nu;\rho)]}, \quad G_{3}(\mu,\nu;\rho):=\dfrac{\Phi_{2}^{\rho}(\mu,\nu;\rho)}{\Phi_{2}(\mu,\nu;\rho)[\Phi(\mu)-\Phi_{2}(\mu,\nu;\rho)]},
\end{align*}
$G_{2,\xi_{0y}}(Z):=G_{2}(Z^{\prime}\mu_{0},X^{\prime}\nu_{y};g(X^{\prime}\rho_{0y}))$, and $G_{3,\xi_{0y}}(Z):=G_{3}(Z^{\prime}\mu_{0},X^{\prime}\nu_{y};g(X^{\prime}\rho_{0y}))$.

The score functions are
\begin{align*}
    S_{10,\xi}(W)&=G_{1}(Z^{\prime}\mu_{0})\left[D-\Phi(Z^{\prime}\mu_{0})\right]Z,\\
    S_{1s,\xi}(W)&=G_{1}(Z^{\prime}\mu_{s})\left[\bar{J}^{s}-\Phi(Z^{\prime}\mu_{s})\right]Z,\\
    S_{2y,\xi}(W)&=D\left\{\bar{I}^{y}\Phi(Z^{\prime}\mu_{0})-\Phi_{2,\xi_{0y}}(Z)\right\} \begin{bmatrix}G_{2,\xi_{0y}}(Z)\\
\dot g(X^{\prime}\rho_{0y})G_{3,\xi_{0y}}(Z)
\end{bmatrix}\otimes X,\\
S_{3sy,\xi}(W)&=D\left\{ \dfrac{\bar{J}^{s}\bar{I}^{y}}{A_1}-\dfrac{\bar{J}^{s}I^{y}}{A_2}-\dfrac{J^{s}\bar{I}^{y}}{A_3}+\dfrac{J^{s}I^{y}}{A_4}\right\} \Phi_{2,\xi_{sy}}^{\rho}(Z)\dot g(Z^{\prime}\rho_{sy}) Z,
\end{align*}
where
\begin{align}\label{eq:As}
A_1 & :=\Phi_{2}(Z^{\prime}\mu_{s},X^{\prime}\nu_{y};g(Z^{\prime}\rho_{sy})),\notag\\
A_2 & :=\Phi_{2}(Z^{\prime}\mu_{s},-X^{\prime}\nu_{y};-g(Z^{\prime}\rho_{sy})),\\
A_3 & :=\Phi_{2}(Z^{\prime}\mu_{0},X^{\prime}\nu_{y};g(X^{\prime}\rho_{0y}))-\Phi_{2}(Z^{\prime}\mu_{s},X^{\prime}\nu_{y};g(Z^{\prime}\rho_{sy})),\notag\\
A_4 & :=\Phi_{2}(Z^{\prime}\mu_{0},-X^{\prime}\nu_{y};-g(X^{\prime}\rho_{0y}))-\Phi_{2}(Z^{\prime}\mu_{s},-X^{\prime}\nu_{y};-g(Z^{\prime}\rho_{sy})).\notag
\end{align}

Let $\partial_{uv} := (\partial^2/\partial u \partial v)$. The components of the expected Hessian functions are:
\begin{footnotesize}
\begin{align*}
    H_{10}&=\Ep\left[\partial_{\mu_{0} \mu_0^{\prime}}\ell_{1,\xi}(W)\right]=-\Ep\left[G_1(Z^{\prime}\mu_0)\dot \Phi(Z^{\prime}\mu_0)ZZ^{\prime }\right],\\
    H_{1s}&=\Ep\left[\partial_{\mu_{s} \mu_s^{\prime}} \ell_{1,\xi}(W)\right]=-\Ep\left[G_1(Z^{\prime}\mu_s)\dot \Phi(Z^{\prime}\mu_s)ZZ^{\prime }\right],
\end{align*}
\[
J_{2y} = \Ep \left[\partial_{\theta_y \mu_{0}'} \ell_{2y,\xi}(W)\right]
= \Ep \left[ \left\{ \dot \Phi(Z' \mu_0) \Phi_{2,\xi_{0y}}(Z)-\Phi(Z^{\prime}\mu_0) \Phi_{2,\xi_{0y}}^{\mu}(Z) 
 \right\} 
\begin{bmatrix} 
G_{2,\xi_{0y}}(Z) \\ 
\dot g(X' \rho_{0y}) G_{3,\xi_{0y}}(Z) 
\end{bmatrix} 
\otimes X Z' \right],
\]
\begin{align*}
H_{2y} =& \Ep\left[\partial_{\theta_y \theta_y'} \ell_{2y,\xi}(W) \right]
\\ &= -\Ep \left[ \Phi(Z^{\prime}\mu_0)
\begin{bmatrix} 
\Phi_{2,\xi_{0y}}^{\nu}(Z) G_{2,\xi_{0y}}(Z) & \Phi_{2,\xi_{0y}}^{\rho}(Z) \dot g(X' \rho_{0y})G_{2,\xi_{0y}}(Z) \\ 
\Phi_{2,\xi_{0y}}^{\nu}(Z) \dot g(X' \rho_{0y}) G_{3,\xi_{0y}}(Z) & \Phi_{2,\xi_{0y}}^{\rho}(Z) \dot g(X' \rho_{0y})^{2} G_{3,\xi_{0y}}(Z) 
\end{bmatrix} 
\otimes X X' \right],
\end{align*}

\[
J_{3sy}^{(1)}= \Ep\left[\partial_{\rho_{sy}  \mu_{0}^{\prime}} \ell_{3sy,\xi}(W)\right]=\Ep\left[\left\{ \dfrac{1}{A_3}\Phi_{2,\xi_{0y}}^{\mu}(Z)-\dfrac{1}{A_4}\left(\dot \Phi(Z^{\prime}\mu_{0})-\Phi_{2,\xi_{0y}}^{\mu}(Z)\right)\right\} \Phi_{2,\xi_{sy}}^{\rho}(Z)\dot g(Z^{\prime}\rho_{sy}) ZZ^{\prime}\right],
\]

\begin{align*}    
J_{3sy}^{(2)} =& \Ep\left[\partial_{\rho_{sy}  \mu_{s}^{\prime}} \ell_{3sy,\xi}(W)\right]\\
&=\Ep\Bigg[\left\{ -\left(\dfrac{1}{A_1}+\dfrac{1}{A_2}+\dfrac{1}{A_3}+\dfrac{1}{A_4}\right)\Phi_{2,\xi_{sy}}^{\mu}(Z)+\left(\dfrac{1}{A_2}+\dfrac{1}{A_4}\right)\dot \Phi(Z^{\prime}\mu_{s})\right\} \Phi_{2,\xi_{sy}}^{\rho}(Z)\dot g(Z^{\prime}\rho_{sy}) ZZ^{\prime}\Bigg],
\end{align*}

\begin{align*}
J_{3sy}^{(3)}=& \Ep\left[\partial_{\rho_{sy} \theta_{y}^{\prime}} \ell_{3sy,\xi}(W)\right]\\
&=\Ep\left[\begin{bmatrix}-\left(\dfrac{1}{A_1}+\dfrac{1}{A_2}+\dfrac{1}{A_3}+\dfrac{1}{A_4}\right)\Phi_{2,\xi_{sy}}^{\nu}(Z)+\left( \dfrac{1}{A_3}+\dfrac{1}{A_4}\right) \Phi_{2,\xi_{0y}}^{\nu}(Z)\\
\left( \dfrac{1}{A_3}+\dfrac{1}{A_4}\right) \Phi_{2,\xi_{0y}}^{\rho}(Z)\dot g(X' \rho_{0y})
\end{bmatrix}\Phi_{2,\xi_{sy}}^{\rho}(Z)\dot g(Z^{\prime}\rho_{sy})\otimes ZX^{\prime}\right],
\end{align*}

\begin{align*}
H_{3sy}=  \Ep\left[\partial_{\rho_{sy}\rho_{sy}^{\prime}} \ell_{3sy,\xi}(W)\right]=\Ep\Bigg[ -\left(\dfrac{1}{A_1}+\dfrac{1}{A_2}+\dfrac{1}{A_3}+\dfrac{1}{A_4}\right)\Phi_{2,\xi_{sy}}^{\rho}(Z)^2\dot g(Z^{\prime}\rho_{sy})^2 ZZ^{\prime}\Bigg].
\end{align*}\end{footnotesize}
Note that $J_{3sy} = \Ep\left[\partial_{\rho_{sy}\eta_{sy}^{\prime}}  \ell_{3sy,\xi}(W)\right] = (J_{3sy}^{(1)},J_{3sy}^{(2)},J_{3sy}^{(3)})$.

By \eqref{eq:fclt},
\begin{align}\label{eq:psi}
\sqrt{n}\left(\hat{\xi}_{sy}-\xi_{sy}\right) =&\sqrt{n}\begin{pmatrix}\hat{\mu}_{0}-\mu_{0}\\
\hat{\mu}_{s}-\mu_{s}\\
\hat{\theta}_{y}-\theta_{y}\\
\hat{\rho}_{sy}-\rho_{sy}
\end{pmatrix} =\dfrac{1}{\sqrt{n}}\sum_{i=1}^{n} \begin{pmatrix}\psi_{10,\xi}(W_i)\\
\psi_{1s,\xi}(W_i))\\
\psi_{2y,\xi}(W_i)\\
\psi_{3sy,\xi}(W_i)
\end{pmatrix} + o_{\Pr}(1) \notag \\& :=  \dfrac{1}{\sqrt{n}} \sum_{i=1}^n  \psi_{sy,\xi}(W_i) + o_{\Pr}(1), \quad  \psi_{sy,\xi}(W_i) :=  - J^{-1}(s,y) \varphi_{sy,\xi}(W_i).
\end{align}
Substituting the expressions of $ \varphi_{sy,\xi}(W_i)$ and $J^{-1}(s,y)$ in  \eqref{eq:scores}  and \eqref{eq:inv-hess} yields $\psi_{10,\xi}(W_i) = -H_{10}^{-1}S_{10,\xi}(W_i)$, $\psi_{1s,\xi}(W_i) = -H_{1s}^{-1}S_{1s,\xi}(W_i)$, $\psi_{2y,\xi}(W_i) = -H_{2y}^{-1}[S_{2y,\xi}(W_i)-J_{2y} H_{10}^{-1}S_{10,\xi}(W_i)]$, and
$$
\psi_{3sy,\xi}(W_i) = -H_{3sy}^{-1} [ S_{3sy,\xi}(W_i)+J_{3sy} \psi_{12sy,\xi}(W_i)],
$$
where $\psi_{12sy,\xi}(W_i) := (\psi_{10,\xi}(W_i)', \psi_{1s,\xi}(W_i)', \psi_{2y,\xi}(W_i)')'$.

We conclude that the asymptotic variance function of $(s,y) \mapsto \hat \rho_{sy}$ is
\begin{equation}\label{eq:avar}
\Sigma_{\rho}(s,y) := \Sigma_{\rho_{sy}\rho_{sy}}=\Ep[\psi_{3sy,\xi}(W_i) \psi_{3sy,\xi}(W_i)^{\prime}].     
\end{equation}

\subsubsection{Estimator of Asymptotic Variance} An estimator of \eqref{eq:avar} can be formed as
\begin{equation}\label{eq:se}
\hat{\Sigma}_{\rho}(s,y)=\dfrac{1}{n}\sum_{i=1}^{n} \hat \psi_{3sy,\hat \xi}(W_i) \hat \psi_{3sy,\hat \xi}(W_i)^{\prime},
\end{equation}
where
\begin{equation}\label{eq:influence}
\hat \psi_{3sy,\xi}(W_i)	=- \hat H_{3sy}^{-1}[S_{3sy,\xi}(W_i)+ \hat J_{3sy} \hat \psi_{12sy,\xi}(W_i) ],
\end{equation}
with
$
\hat H_{3sy} = n^{-1} \sum_{i=1}^{n} \partial_{\rho_{sy} \rho_{sy}^{\prime}} \ell_{3sy,\hat \xi}(W_i), 
$
$
\hat J_{3sy} = n^{-1} \sum_{i=1}^{n}  \partial_{\rho_{sy}\eta_{sy}^{\prime}}  \ell_{3sy,\hat \xi}(W_i),
$
\begin{equation}\label{eq:influence12}
\hat \psi_{12sy,\xi}(W_i) = - \begin{pmatrix} \hat H_{10}^{-1}S_{10,\xi}(W_i)\\
 \hat H_{1s}^{-1} S_{1s,\xi}(W_i)\\
 \hat H_{2y}^{-1} [S_{2y,\xi}(W_i)- \hat J_{2y} \hat H_{10}^{-1}S_{10,\xi}(W_i)]
\end{pmatrix},
\end{equation}
$
\hat H_{10}	= n^{-1} \sum_{i=1}^{n} \partial_{\mu_{0} \mu_0^{\prime}}\ell_{10,\hat \xi}(W_i)$, $
\hat H_{1s}	= n^{-1} \sum_{i=1}^{n} \partial_{\mu_{s} \mu_s^{\prime}}\ell_{1s,\hat \xi}(W_i)$, $
\hat H_{2y}	= n^{-1} \sum_{i=1}^{n} \partial_{\theta_y \theta_y'} \ell_{2y,\hat \xi}(W_i) $ and $
\hat J_{2y} = $ $n^{-1} \sum_{i=1}^{n} \partial_{\theta_y \mu_{0}'} \ell_{2y,\hat \xi}(W_i)$. In the previous expressions we use the notation $$f_{\hat \xi}(W_i) :=  f_{\xi}(W_i) \Big|_{\xi = \hat \xi}.$$ For example, $\hat \psi_{3sy,\hat \xi}(W_i)$ is equal to $\hat \psi_{3sy,\xi}(W_i)$ evaluated at $\xi = \hat \xi$.

\end{document}